\documentclass[conference]{IEEEtran}

\IEEEoverridecommandlockouts
\usepackage{cite}
\usepackage{algorithm}
\usepackage{amsmath,amssymb,amsfonts}
\usepackage{algorithmicx}
\usepackage{algpseudocode}
\usepackage{graphicx}
\usepackage{textcomp}
\usepackage{xcolor}
\usepackage{multirow}
\usepackage{latexsym}
\usepackage{subfigure}
\usepackage{tabularx}
\usepackage{booktabs}
\usepackage{longtable}
\usepackage{diagbox}
\usepackage{enumitem}
\usepackage{footnote}

\usepackage[affil-it]{authblk}
\usepackage[english]{babel}
\usepackage{blindtext}

\usepackage[T1]{fontenc}
\usepackage[utf8]{inputenc}
\usepackage{authblk}
\usepackage{textcomp}

\usepackage[affil-it]{authblk}
\usepackage[english]{babel}
\usepackage{blindtext}

\begin{document}
\title{Mitigating Popularity Bias in Recommendation with
Unbalanced Interactions: A Gradient Perspective
}

\iffalse
\author{\IEEEauthorblockN{Given Name Surname \footnotemark[1]}
\IEEEauthorblockA{\textit{dept. name of organization (of Aff.)} \\
\textit{name of organization (of Aff.)}\\
City, Country \\
email address or ORCID}
\and
\IEEEauthorblockN{Given Name Surname \footnotemark[1]}
\IEEEauthorblockA{\textit{dept. name of organization (of Aff.)} \\
\textit{name of organization (of Aff.)}\\
City, Country \\
email address or ORCID}
\and
\IEEEauthorblockN{3\textsuperscript{rd} Given Name Surname}
\IEEEauthorblockA{\textit{dept. name of organization (of Aff.)} \\
\textit{name of organization (of Aff.)}\\
City, Country \\
email address or ORCID}
\and
\IEEEauthorblockN{4\textsuperscript{th} Given Name Surname}
\IEEEauthorblockA{\textit{dept. name of organization (of Aff.)} \\
\textit{name of organization (of Aff.)}\\
City, Country \\
email address or ORCID}
\and
\IEEEauthorblockN{5\textsuperscript{th} Given Name Surname \footnotemark[1] }
\IEEEauthorblockA{\textit{dept. name of organization (of Aff.)} \\
\textit{name of organization (of Aff.)}\\
City, Country \\
email address or ORCID}
\and
\IEEEauthorblockN{Given Name Surname \footnotemark[1]}
\IEEEauthorblockA{\textit{dept. name of organization (of Aff.)} \\
\textit{name of organization (of Aff.)}\\
City, Country \\
email address or ORCID}
}
\fi

\author[*$\sharp$]{Weijieying Ren \thanks{* denotes equal contribution.}}
\author[*$\flat$]{Lei Wang} 
\author[$\#$]{Kunpeng Liu}
\author[$\S$]{Ruocheng Guo}
\author[$\flat$ $\textleaf$]{LIM Ee Peng\thanks{$\textleaf$ denotes corresponding author.}}
\author[$\sharp$ $\textleaf$]{Yanjie Fu}
\affil[$\sharp$]{Department of Computer Science, University of Central Florida, USA}
\affil[$\flat$]{Department of Computer Science, Singapore Management University, Singapore}
\affil[$\#$]{Department of Computer Science, Portland State University, USA}
\affil[$\#$]{ByteDance AI lab, UK}
\affil[$\null$]{\textit {wjyren@knights.ucf.edu, lei.wang.2019@phdcs.smu.edu.sg, kunpeng@pdx.edu }}
\affil[$\null$]{\textit {ruocheng.guo@bytedance.com, eplim@smu.edu.sg, yanjie.fu@ucf.edu}}
\renewcommand\Authands{ and }

\maketitle
%\footnotetext[1]{Corresponding author}
%\footnotetext[*]{indicates equal contribution}

\begin{abstract}
Recommender systems learn from historical user-item interactions to identify preferred items for target users. These observed interactions are usually unbalanced following a long-tailed distribution. Such long-tailed data lead to popularity bias to recommend popular but not personalized items to users. 
We present a gradient perspective to understand two negative impacts of popularity bias in  recommendation model optimization: (i) 
the gradient direction of popular item embeddings is closer to that of positive interactions, and (ii) the magnitude of positive gradient for popular items are much greater than that of unpopular items. 
To address these issues, we propose a simple yet efficient framework to mitigate popularity bias from a gradient perspective. 
Specifically, we first normalize each user embedding and record accumulated gradients of users and items via popularity bias measures in  model training. 
To address the popularity bias issues, we develop a  gradient-based embedding adjustment approach used in model testing. 
This strategy is generic,  model-agnostic, and can be seamlessly integrated into most existing recommender systems.
Our extensive experiments on two classic recommendation models and four real-world datasets demonstrate the effectiveness of our method over state-of-the-art debiasing baselines.

\end{abstract}

% \vspace{-10px}
\section{Introduction}
Recommender systems (RS) provide users with newly suggested items by mining user preferences from historical user-item interactions \cite{yang2018unbiased,xu2018enhancing,ren2017robust,chen2020bias}.
% As online shopping and advertising applications become widespread, there is an emerging RS research trend on using implicit user feedback (e.g., purchases, clicks, browses) to recommend future items. 
Each observed user-item interaction represents an implicit feedback (e.g., purchases, clicks, browses) indicating the user’s preference on the item.  We call this a \textit{positive} interaction.  To train a RS model, we can uniformly select a few items having no interaction with the user as \textit{negative} items. A RS model based on such implicit feedback data is designed to predict  positive items with higher prediction scores than negative items. Generally, each item will receive a \textit{positive gradient} if it has observed interactions with a user and obtain a \textit{negative gradient} if it is selected as a negative sample.

Nevertheless, recent studies~\cite{yang2018unbiased,chen2020bias,wang2022explanation} found that in most real-world application scenarios, user-item interactions in recommender systems are highly imbalanced following a long-tailed distribution~\cite{yin2020learning,lee2011my}. Only a small number of items are popular as they frequently interact with users and thus are more likely favored by users. 
%The interactions between popular items and users thus account for a large number of user-item interactions. 
When such long-tailed data are used for training a RS model, \textit{popularity bias} arises in the trained model, i.e, popular items are overly recommended to all users, while some less popular items may be more relevant to some specific users.
Popularity bias can result in low visibility of unpopular items, reduced diversity of markets and damage consumer experience. Mitigating popularity bias is therefore highly demanded. 
%, and most importantly unfairness. Unfairness here refers to recommendation outcome not consistent with the true preference of users.  
\begin{figure}[t!] 
\centering
\includegraphics[width=1.0\linewidth]{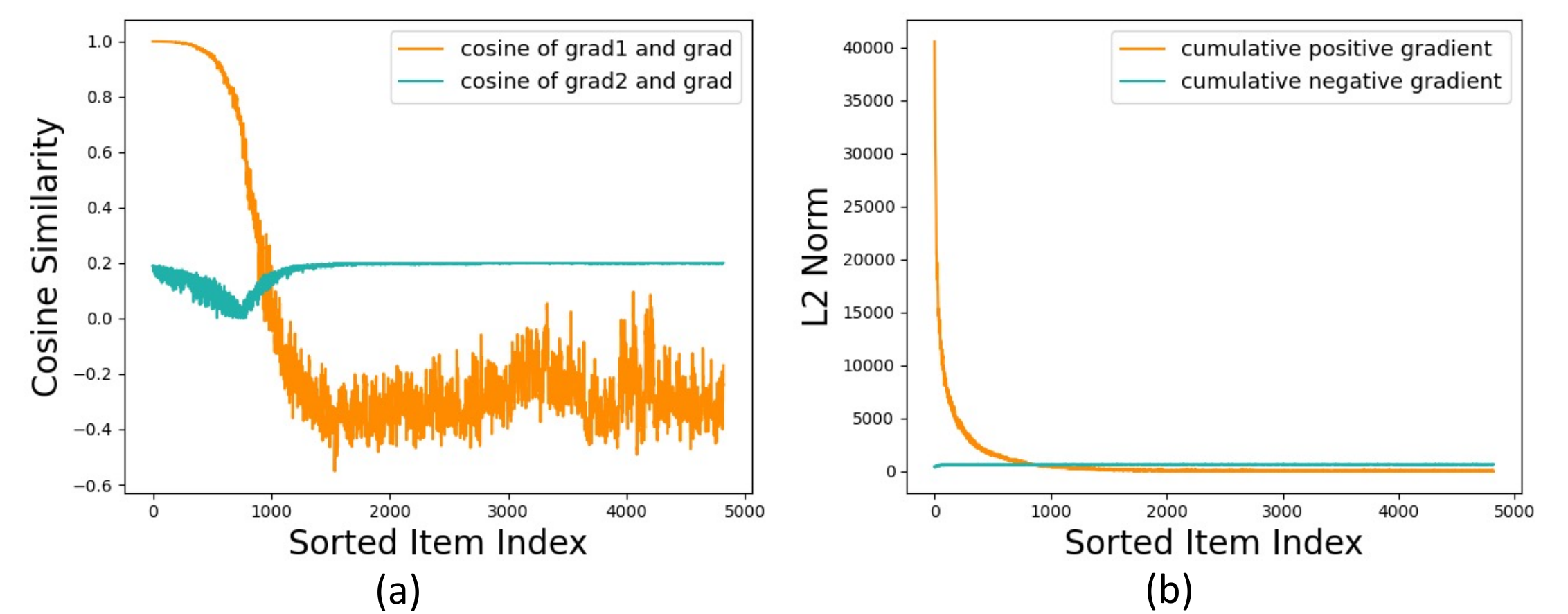} 
\vspace{-15px}
\caption{
(a) Directional similarity between accumulated positive gradient (negative gradient) and combined gradient of item embeddings shown as orange dots (green dots) for the Movielens-10M dataset. We define directional similarity by cosine similarity function.  Note that items are sorted in inverse popularity order, i.e., more popular items are assigned smaller index on the $x$-axis. (b) Magnitude (L2 norm) of accumulated positive gradient (negative gradient) of item embeddings shown as orange dots (green dots) for items sorted in inverse popularity order. 
% \zxy{enlarge all text}
% Comparison of L2 norm values of the positive and negative gradients of items on Movielens-10M dataset.
}
\vspace{-15px}
\label{Fig.main1} 
\end{figure}
%By Yanjie 
%paragraph2
%research gap of existing work
% literature 1 and weaknesses
% literature 2 and weaknesses

To deal with popularity bias, one prominent direction is to design a re-weighting loss, e.g., inverse propensity score (IPS), to balance overall item distributions \cite{austin2015moving,ge2021towards}. Another alternative is to conduct knowledge transfer \cite{yin2020learning} with the assumption that knowledge represented in popular items should be beneficial to the user preference mining in less popular items. Recently, more sophisticated models are designed to preserve user specific tastes via the usage of side information \cite{li2021leave}, autoencoder model \cite{wei2020model} and to disentangle user preference following pre-defined prior heuristics \cite{zheng2021disentangling}.
However, previous works either require complex model design \cite{wei2020model}, degenerate embedding representation or rely on strong assumption \cite{zheng2021disentangling}, which is lack of generalization ability.
%\zxy{what is the disadvantages of these methoeds?}
% {\it (3) Utility fairness enforcement based strategy}, which enforces popular and unpopular items to receive fair engagement metrics proportional to their true audience sizes \cite{ge2021towards,xiao2017fairness,fu2020fairness}. However, a low-quality item may not deserve 
%{\color{blue} (EP: be assigned?)}
% a high engagement metric for it to be unpopular. 
% Therefore, enforcing such irrational fairness may unwittingly introduce another type of bias and confusion to the recommendation model.  
%Also, disentangle representation based methods, which leverage disentangle representation learning methods to separate the embedding of item true characteristics and item popularity bias~\cite{???}
\begin{figure}[t!] 
\centering
\includegraphics[width=0.8\linewidth]{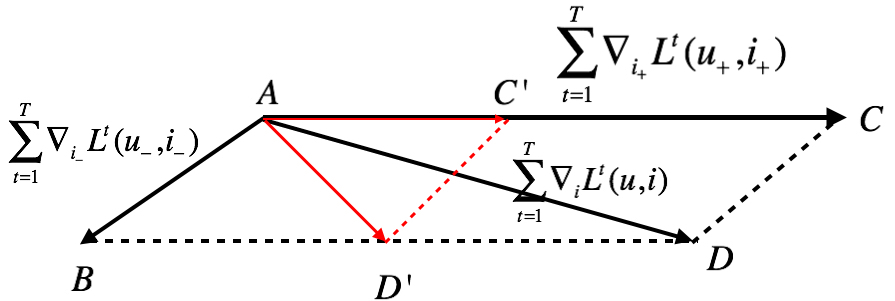} 
\vspace{-10px}
\caption{For a popular item $i$, let $\protect\overrightarrow{AC}$ be its accumulated positive gradient and $\protect\overrightarrow{AB}$ denotes the accumulated negative gradient. The overall gradient (learned vector) is thus $\protect\overrightarrow{AD}=\protect\overrightarrow{AB}+\protect\overrightarrow{AC}$ which is still very similar to $\protect\overrightarrow{AC}$. Our gradient based embedding adjustment approach is to reduce $\protect\overrightarrow{AC}$ to $\protect\overrightarrow{AC'}$ leading to a new overall gradient $\protect\overrightarrow{AD'}$ which effectively incorporates the negative gradient.}
\vspace{-10px}
\label{Fig.main2} 
\end{figure}
Unlike the aforementioned strategies, we approach the popularity bias by analyzing gradient distortion in long-tailed data, attributed to the combination of gradients generated by positive and negative items.  We discover that (\textbf{\textit{i}}) the overall gradient direction of popular items' embeddings tend to be closer to the gradient of positive interactions, and (\textbf{\textit{ii}}) the magnitude of positive gradient for popular items 
is far larger than that of unpopular items.

To illustrate the above challenges, we visualize the related variables in the training process of Movielens-10M dataset in Fig.\ref{Fig.main1}. Specifically, the cosine similarity between the accumulated positive gradient (given by $\sum_{t=1}^{T}\bigtriangledown_i \mathcal{L}^t(u_+,i_+)$, where $t$ is the iteration step) and overall gradient of each item is visualized in Fig.~\ref{Fig.main1}(a) for items sorted by inverse popularity order.  We also show the cosine similarity between the
% {\color{red} accumulated (EP: need definition)} 
negative gradient and overall gradient of items.  The high similarity (i.e., close to 1) between the overall gradient and the accumulated positive gradient of popular items indicates that the overall gradient of each popular item is shifted towards the direction of its positive interactions.  The opposite can be observed for unpopular items.  Fig.~\ref{Fig.main1}(b) shows the norm of accumulated positive and negative gradients for items sorted in the descending popularity order. In this figure, popular items exhibit large accumulated positive gradient norm values than their accumulated negative gradient norm values.  In other words, the accumulated gradients of popular items increase the norm of their corresponding item embedding vectors and results in tendency to rank high in recommendation results.

To further analyze the underlying reasons of the above empirical findings, we review the optimization process of the classical pairwise ranking loss used in implicit RS, i.e., BPR loss \cite{rendle2012bpr}.
%Ignoring the effect of hyper-parameters, e.g., learning rate, 
It is clear that item (or user) embedding vectors are mainly determined by their accumulated gradients. 
%As shown in Fig.2, for a specific item $i$, line AC represents the accumulated positive gradients in previous optimization stage and line AB denotes the negative 
Considering a popular item $i$ illustrated in Fig.~\ref{Fig.main2}, 
vectors $\overrightarrow{AC}$ and $\overrightarrow{AB}$ represent its accumulated positive and negative gradient respectively. The overall gradients $\overrightarrow{AD}$ is shifted towards $\overrightarrow{AC}$ due to the latter's large magnitude, making the learned item embedding only facilitate the learning of positive interaction (vice versa for unpopular items). The accumulated gradients are derived by summing up gradient values across all iterations when optimizing the objective function $\mathcal{L}^t(u,i)$. Consequently, 
We identify three factors which may account for the reason of the existence of popularity bias in implicit RS: 1) item popularity, %Popular items are exposed more in loss optimization process. 
2) user activity, 
%The iterative optimization of user/item embedding makes their information are mutually influenced. An active user contain large embedding magnitude and contribute more to item gradients. 
and 3) user conformity.
% (see Section 3).
%If a user follows the main taste of other users, their corresponding items will be updated more. 

\textbf{Our Approach.}
Motivated by mitigating popularity bias, we found that the combined analysis for gradients generated by positive and negative interactions are key solutions. Hence, we propose the concept of user/item embedding adjustment. In the training stage, we only normalize user embeddings to mitigate the influence of high activity users and record the accumulated gradients of users and items. 
In the testing stage, we propose to mitigate bias by adjusting the learned user and item embeddings to mitigate the influence by item popularity and user conformity.
% Motivated by this, we hypothesize that the combined analysis for gradients generated by positive and negative interactions could be improper to mitigate popularity bias and attempt to modified the learned item (user) embedding.  We thus propose the conception of user/item embedding disentanglement. In the training stage, we only normalize user embedding to mitigate the influence of user activity. In the test stage, as demonstrated in Fig.2, we modify the learned item (user) embedding from AD to an unbiased embedding AD`, which could be benefit of both the positive and negative interaction. To be specific, we disentangle item popularity (user popularity) from learned item (user) embedding.  
%Our proposed method is easily to be implemented and is applicable to most existing recommendation systems.
%textcolor{blue}{delete??? However, it is non-trivial to implement the accumulated gradient surgery since 1) the iterative optimization of user-item embedding makes it unclear how much item popularity and user conformity contributes to the biased overall gradient; 2) it is of great challenge to define or even formulate the optimal direction and magnitude of unbiased accumulated overall gradient AD`. To simplify heuristic, in this paper, we modify the biased accumulated gradient AD through its accumulated positive gradients AC. The modifies AC' (length of AC should be larger than AC for unpopular items) along with AB. {\color{blue}lei: to be discussed.} } 
%In summary, our main contributions are as follows:
Our main contributions are as follows:
% \vspace{-5px}
\begin{itemize}[leftmargin=*]
\item \textbf{A Novel Gradient Perspective to Understand Popularity Bias Mechanism.} Classical methods focus on understanding popularity bias from balancing the overall data distribution, e.g., popular and unpopular groups. In this paper, we provide new findings that 1) the overall gradient direction of popular (unpopular) items is shifted to be closer to the gradient on positive (negative) interactions; 2) positive gradient magnitude of popular items is usually much larger than that of unpopular items. We analyze such phenomenon from the gradient perspective, and provide a new insight on balancing the positive and negative gradients for the same item.
\item \textbf{Gradient Based Embedding Adjustment Approach.}
Motivated by the new findings, we identify three essential factors, namely item popularity, user conformity, and user activeness and analyze how user and item embeddings are related to them during the optimization steps. We develop a simple yet novel embedding post-doc adjustment method. This framework is explainable and interpretable.
% (See Section 4). 
\item \textbf{Extensive Experiments with Real-world Datasets.} We conduct extensive experiments 
% on four real-world datasets 
to evaluate the effectiveness of our proposed methods in terms of overall accuracy, generalisation, and efficiency against state-of-the-art baselines. 
% The experimental results demonstrate that our approach outperform the start-of-the models (i.e., DICE~\cite{zheng2021disentangling} and MACR~\cite{wei2020model}) 
We also conducted ablation studies to evaluate the components of our proposed methods.
\end{itemize}

\section{Preliminaries}
\textbf{Problem Statement.}  Suppose we have a user set $\mathcal{U}$, an item set $\mathcal{I}$ and an user-item interaction matrix $ Y \in \{0,1\} ^{|\mathcal{U}| \times |\mathcal{I}|}$, where $Y_{ui}$ indicates whether $u \in \mathcal{U}$ has made an interaction (e.g., adoption or purchase) with item $i \in \mathcal{I}$ or not. 
% \blue{to be refined.}
The goal of a recommender is to learn a scoring function $r(u,i)$ to recommend a list of $k$ items to a user $u$:
%based on the user's implicit feedback matrix $Y$:
\begin{equation}
    \hat{y}_{ui} = r(u,i) = \boldsymbol{P_u}^{T}\boldsymbol{Q_i}
\end{equation}
% The goal of a recommender is to learn a scoring function $r(u,i|\Theta)$ with parameters $\Theta$ to recommend a list of $k$ items to a user $u$ based on the user's implicit feedback matrix $Y$:

% \begin{equation}
    % \hat{y}_{ui} = r(u,i|\boldsymbol{P_u},\boldsymbol{Q_i}) = \boldsymbol{P_u}^{T}\boldsymbol{Q_i}
% \end{equation}
where $\hat{y}_{ui}$ is the predicted score, $\boldsymbol{P_u}$ is the user representation vector for the user $u$, and $\boldsymbol{Q_i}$ is the item representation vector for the item $i$. $\boldsymbol{P_u}$ ($\boldsymbol{Q_i}$) is derived from the user encoder $f(u|\Theta_U)$ (the item encoder $f(i|\Theta_I)$). Note that the user or item encoder is agnostic to model, which can be the embedding matrix~\cite{koren2009matrix}, graph neural network~\cite{he2020lightgcn}, or other models. For simplicity, we refer to $\boldsymbol{P_u}$ ($\boldsymbol{Q_i}$) as user embedding (item embedding) in the rest of the paper. 
% user’s embedding is denoted as $\boldsymbol{P_u}$ and item embedding is marked as $\boldsymbol{Q_i}$, respectively, 

% {\color{red} what is $\hat{y}_{ui}$ here? $Y_{ui}$ no elaboration}

\iffalse
%In traditional train-test data split such as random split and split-by-time, training and test sets follow the same or similar distribution. However, real-world recommender systems are often trained and updated from time to time with evolving user interactions, i.e., non-IID data. To recommend for a user based on true and unbiased user preference, we %prepare data for performing unbiased learning from biased data. Specifically, we 
let training data keep the popularity bias and make the test data capture a uniform distribution over items.

\ParaTitle{Scoring function} $r(u,i|\Theta)$ estimates the relevance between a user $u$ and item $i$ based on user’s embedding $\boldsymbol{P_u}$ and item’s embedding $\boldsymbol{Q_i}$. As a widely adopted recommendation method, matrix factorization \cite{koren2009matrix} formulates the scoring function as follows:
\fi

\textbf{Why does the embedding magnitude matter in RS problems?} In the testing stage, the recommender system returns the personalized ranking of items $R_u$ for a user $\boldsymbol{u}$:  
\begin{equation}
\begin{aligned}
    R_u &= \{ (i,rank(\hat{y}_{ui}) \}_{i \in I}  \\ 
    &= \{ (i,rank(\boldsymbol{P_u}^{T}\boldsymbol{Q_i})) \}_{i \in I}\\
    &= \{ (i,rank(\| \boldsymbol{Q_i} \| cos(\boldsymbol{P_u},\boldsymbol{Q_i}))) \}_{i \in I}
    \label{eq.ranking}
\end{aligned}
\end{equation}
where $rank$ is a ranking function based on a set of scoring function values between $u$ and all items $i \in I$. In Eq.~\ref{eq.ranking}, we omit $\| \boldsymbol{P_u} \|$ as $R_u$ involves the same user $u$.
In other words, the personalized ranking result during the testing stage only depends on the magnitude of item embedding $\boldsymbol{Q_i}$ and its cosine similarity with the user embedding $\boldsymbol{P_u}$. 

\textbf{How are the embedding magnitudes generated in RS?} In the training stage, we often adopt a pairwise ranking loss, e.g., Bayesian Personalized Ranking (BPR) loss \cite{rendle2012bpr} and Binary cross entropy (BCE) \cite{xue2017deep} loss. Without loss of generality, we revisit BPR loss functions for recommender system with implicit feedback. 
%The analytical results shed light on the root cause (Section 3) that affects user and item embeddings and inspire us to design a simple yet effective solution (see Section 4). 
%~\ref{sec:approach}.
% \vspace{5pt}
% \ParaTitle{Iterative Update Rule.}
In realistic scenarios, only positive implicit feedback (e.g., clicks and purchases) are observed. 
In implicit RS, we assume that user $u$'s preference over an item $i$ with an observed $(u,i)$ interaction is ranked higher than the unobserved $(u,j)$ interaction as a negative sample, 
%$ i \succ_uj
%\label{eq::ranking}
%$,
where the item $j$ is uniformly selected from item set $I$ that does not have observed interactions with $u$. Without loss of generality, we represent BPR loss here to maximize the likelihood of observing such pairwise ranking relations:
\begin{equation}
    L_{BPR} = \sum_{(u,i,j)\in D_s }  \log \;(\sigma(\hat{y}_{ui}-\hat{y}_{uj}))-\lambda_\Theta ||\Theta||^2
\end{equation}
where $D_s$ is a set of $(u,i,j)$ triplets with $(u,i)$ and $(u,j)$ as positive and negative samples respectively. $\Theta$ is the model parameters and $\sigma(x) = \frac{1}{1+\exp(-x)}$ is the sigmoid function. %Formally, the problem of predicting $\hat(r_{ui})$ can be seen as the task of estimating a matrix $R : U \times I$. Thus, this prediction formula can be rewritten as:
%\begin{equation}
%    \hat{r_{ui}} = <w,h> = \sum_{f=1}^k w_{uf} \cdot h_{if} 
%\end{equation}
The gradient of BPR loss with respect to the model parameters $\Theta$ (e.g., user/item embeddings) can be written as:
\begin{equation}
\begin{aligned}
    \frac{\partial \; L_{BPR}}{\partial\; \Theta}&=\sum_{(u,i,j) \in D_S} \frac{\partial \, \log \sigma(\hat{y}_{ui}-\hat{y}_{uj}))}{\partial \, \Theta} - \lambda_{\Theta} 
\frac{\partial \, ||\Theta||^2}{\partial \, \Theta } \\
& \propto \frac{-e^{-(\hat{y}_{ui}-\hat{y}_{uj})}}{1+-e^{-(\hat{y}_{ui}-\hat{y}_{uj})}} \cdot \frac{\partial \,(\hat{y}_{ui}-\hat{y}_{uj})}{\partial \, \Theta} - \lambda_{\Theta} \Theta
\end{aligned}
\end{equation}

%Considering the partially interaction records, \cite{} Personalized Bayesian Ranking (BPR) loss is proposed to model the ranking order than optimizing prediction loss on the whole item lists. \cite{}. 
More concretely, the gradients for updating a matrix factorization model are:
\begin{equation}
\frac{\partial \; (\hat{y}_{ui}-\hat{y}_{uj})}{\partial \; \Theta}
=
\left\{\begin{matrix}
 (\boldsymbol{Q_{i}}-\boldsymbol{Q_{j}}) & \; \text{if} \; \Theta = \boldsymbol{P_{u}} \\
 \boldsymbol{P_{u}} & \; \text{if}  \; \Theta = \boldsymbol{Q_{i}}\\ 
 - \boldsymbol{P_{u}} & \; \text{if}  \; \Theta = \boldsymbol{Q_{j}}\\ 
\boldsymbol{0} & \; \text{otherwise}
\label{eq.gradient}
\end{matrix}\right.
\end{equation}

\textbf{Failure case analysis}. For BPR loss optimization, we notice that for each user-item pair $(u,i)$ with positive interaction, the model will randomly sample a negative pair $(u,j)$. For a popular item $i$ which matches mainstream preference, it is frequently updated by a positive gradient $\boldsymbol{P_{u}}$. It is only updated by a negative gradient $-\boldsymbol{P_{u}}$  when it serves as a negative sample.
In other words, item $i$ will be updated with positive gradients more frequently than negative gradients during the optimization. 
%We discuss it in detail in Section~\ref{sec:motivation}.
% The cumulative gradient update has a non-negligible impact on the optimized model, which can verified in Figure~\ref{Fig.main1}.

\section{Motivation: Popularity Bias from a Gradient Perspective}
\label{sec:motivation}
In this section, we discuss popularity bias from a gradient update perspective, i.e., how the item popularity, user activity and user conformity influence the gradient magnitude and shift the overall gradient to the positive interaction direction for popular items.

Based on Eq.\ref{eq.gradient}, we can derive that user (item) embeddings affect the magnitude and direction of the item (user) embeddings. 
For simplicity, we define the update of the latent vector  of positive items $\boldsymbol{Q^{t}_i}$, negative items $\boldsymbol{Q^{t}_j}$, and users $\boldsymbol{P^{t}_u}$ in the $t^{th}$ iteration step for a data example $(u,i,j)$ with the learning rate $lr$ as follows:
\begin{equation}\label{eq.7}
\begin{aligned}
\boldsymbol{Q^t_i} &=  \boldsymbol{Q^{t-1}_i} - lr\cdot \boldsymbol{P^{t-1}_u},  \\
\boldsymbol{Q^t_j} &=  \boldsymbol{Q^{t-1}_j} - lr\cdot (-\boldsymbol{P^{t-1}_u}),\\
\boldsymbol{P^t_u} &=  \boldsymbol{P^{t-1}_u} - lr\cdot (\boldsymbol{Q^{t-1}_i}-\boldsymbol{Q^{t-1}_j})
\end{aligned} 
\end{equation}

\begin{table}[t]
\caption{Statistics of the training data in our Movielens-10M dataset. $\mathrm{pop(unp)\_i4u}$ and $\mathrm{act(ina)\_u4i}$ indicate the average number of popular (unpopular) items for a user and active (inactive) users for an item respectively.
% We sort users by the number of interactions in descending order and regard sorted users whose interactions accounted for 80\% as active users. In the same way, we can get popular items.
}
% $\mathrm{act(ina)\_u4i}$ denotes the average number of interacted active(inactive) users for an item. $+$ and $-$ represents user-item interaction and negative samples, respectively.}

% We sort users by the number of interactions in descending order and regard sorted users whose interactions accounted for 80\% as active users. In the same way, we can get popular items.

\vspace{-7px}
\centering
% \vspace{-5px}
\resizebox{0.34\textwidth}{!}{
{\small
\begin{tabular}{l@{\hskip 0.2in}c@{\hskip 0.2in}cc}
% \begin{tabular}{lccc}
\toprule
User Group & \#Users &  $\mathrm{pop\_i4u}$ &  $\mathrm{unp\_i4u}$ \\
% $\mathrm{pop^-\_i4u}$ & $\mathrm{unp^-\_i4u}$\\
\midrule
All  & 37,962 & 17.3 & 4.3\\ 
% &1.5 & 20.1\\
Active & 20,801 & 25.1 & 6.6\\ 
% & 2.2 & 29.5\\
Inactive & 17,161 & 7.9 & 1.7 \\ 
% & 0.6 & 8.9\\
\midrule
\midrule
Item Group & \#Items &
$\mathrm{act\_u4i}$ &  $\mathrm{ina\_u4i}$ \\
% $\mathrm{act^-\_u4i}$ & $\mathrm{ina^-\_u4i}$\\
\midrule
All & 4,819 & 136.6 & 34.2 \\
% & 77.2 & 93.6\\
Popular & 328 & 1,591.6 & 414.0 \\
% &1,098.9 & 906.5 \\
Unpopular & 4,491 & 30.3 & 6.4 \\
% & 16.6 & 20.1\\

\bottomrule
\end{tabular}\label{tab:v}
}}
\vspace{-10px}
\end{table}
For supporting the analysis of popularity bias from a gradient perspective. We take the Movielens-10M as an example, and statistics is shown in Table~\ref{tab:v}.
We sort items by their number of interactions in descending order and select sorted items until the number of interactions of selected items up to 80\% of total interactions in the training set (in fact, less than 20\% items account for more than 80\% interactions in many real scenarios). We assume this selected items as popular items and the remaining items as unpopular items. In the same way, we can get active users and inactive users.

\textbf{Influence of User Activity on Item Side.}
Let the latent vector of item $i$ in the $0$-th iteration be $\boldsymbol{Q^0_i}$. Then we can obtain the equivalent form of $\boldsymbol{Q^t_i}$ as:
\begin{equation}\label{eq.8}
\boldsymbol{Q^t_i} =  \boldsymbol{Q^{0}_i} - lr\cdot (\sum_{(u,i)\in Y^{+,t-1}_{i}}\boldsymbol{P_{u}} - \sum_{(u',i)\in Y^{-,t-1}_{i}}\boldsymbol{P_{u'}} )
\end{equation}
where $Y^{+,t-1}_{i}$ and $Y^{-,t-1}_{i}$ denote the set of use-item pairs in which the item $i$ works as the positive and negative items in training set before the iteration step $t$, respectively. Without loss of generalization, we simplify all equations later by ignoring iteration steps $t$ and $t-1$. We let $\Delta_i$ represent the increment of $\boldsymbol{Q^t_i}$ :   
\begin{equation}\label{eq.9}
\Delta_i = \sum_{(u,i)\in Y^{+}_{i}}\boldsymbol{P_{u}} - \sum_{(u',i)\in Y^{-}_{i}}\boldsymbol{P_{u'}} 
\end{equation}
Since users in both $Y^{+}_{i}$ and $Y^{-}_{i}$ consist of active users and inactive users, equation \ref{eq.9} can be extended as:

\begin{equation}\label{eq.10}
\begin{aligned}
\Delta_i &=\sum_{(u,i)\in Y^{+}_{i, act}}\boldsymbol{P_{u}} + 
\sum_{(u,i)\in Y^{+}_{i, inact}}\boldsymbol{P_{u}} \\
&- \sum_{(u',i)\in Y^{-}_{ i,act}}\boldsymbol{P_{u'}} - 
\sum_{(u',i)\in Y^{-}_{ i, inact}}\boldsymbol{P_{u'}}
\end{aligned}
\end{equation}
where $Y^{+}_{i,act}$ ($Y^{+}_{i,inact}$) is the set of use-item pairs for item $i$ in which users are active (inactive) users.

As shown in Table~\ref{tab:v}, for items in any group (all, popular, and unpopular), the number of positive active users is much larger than inactive users.
% ~\footnote{We show detailed statistics about item groups in Movielens-10M in Appendix A.1.}.  
Besides, the number of negative active users is much smaller than inactive users via negative sampling over uniform distribution. Thus, the items embeddings are easily dominated by active users Inferential Failure, according to the cumulative gradients calculated by Eq.~\ref{eq.9}. 
% As shown in Fig.\ref{fig:analysis_1} (a), 
We can derive that the higher the popularity of the item is, the larger the magnitude of its corresponding embedding will be. Moreover, the distance between popular items might be close in latent space due to the user conformity.
% \footnote{We show an illustration of the magnitude of item embeddings in Appendix.}.

% \ParaTitle{Gradient Perspective on User Side.}
\textbf{Influence of Item Popularity on User Side.} Similar to analyze items from the gradient perspective, we can obtain accumulated gradients $\Delta_u$ of user $u$ by:
\begin{equation}
\begin{aligned}
\Delta_u &=\sum_{(u,i)\in Y^{+}_{u,pop}}\boldsymbol{Q_{i}} + 
\sum_{(u,i)\in Y^{+}_{u, unp}}\boldsymbol{Q_{i}} \\
&- \sum_{(u,j)\in Y^{-}_{u, pop}}\boldsymbol{Q_{j}} - 
\sum_{(u,j)\in Y^{-}_{ u,unp}}\boldsymbol{Q_{j}}
\label{eq::last}
\end{aligned}
\end{equation}
% we let the latent vector of user $u$ in the $0$-th iteration be $P^0_u$, and then we can define $P^t_u$ as:
% \begin{equation}\label{eq.11}
% \boldsymbol{P^t_u} =  \boldsymbol{P^0_u} - \textcolor{red}{lr}\cdot \sum_{(u,i,j)\in D_s}(\boldsymbol{Q^{t-1}_i}-\boldsymbol{Q^{t-1}_j}).
% \end{equation}

% For simplicity, we refer to Eq.~10 and can obtain accumulated gradients $\Delta_u$ of user $u$ as:
% \begin{equation}\label{eq.8}
% \Delta_u =\sum_{(u,i)\in Y^{+}_{u,pop}}\boldsymbol{Q_{i}} + 
% \sum_{(u,i)\in Y^{+}_{u, unp}}\boldsymbol{Q_{i}} - 
% \sum_{(u,j)\in Y^{-}_{u, pop}}\boldsymbol{Q_{j}} - 
% \sum_{(u,j)\in Y^{-}_{ u,unp}}\boldsymbol{Q_{j}}, 
% \label{eq::last}
% \end{equation}
For users in any group (all, active, inactive), the number of positive popular items is much larger than unpopular items
% \footnote{We show detailed statistics about user groups in Movielens-10M in Appendix A.1.}.
The number of negative popular items is however much smaller than unpopular items in randomly negative sampling. Therefore, popular items significantly dominate user embeddings due to the accumulated gradients defined in Eq.~\ref{eq::last}. 

\section{Proposed Solution}
\label{sec:approach}
%In the previous section, we have analyze that optimization of BPR loss will assign different proportions of positive and negative gradients for each item. This explains why popular items is overly recommended and unpopular items are often ignored.
%
In this section, we present our method. Specifically, in the training stage, our method normalizes only user embeddings after learning the user and item embeddings using {LGN\cite{he2020lightgcn} or MF \cite{koren2009matrix}}. We then apply post-hoc debiasing on user and item embeddings in the testing stage to mitigate item popularity bias. We first give our definition of an unbiased item (user) embedding in implicit RS. We next give a detailed description about how to mitigate user activity in the training stage, and how to mitigate item popularity and user conformity influence.
\begin{itemize}[leftmargin=*]
    \item \textbf{Disentangled embedding learning}. We first define four factors influencing user and item embedding learning.
    \item \textbf{A Post-hoc debiasing method.} Following the gradient optimization process, we analyze how the four factors propagate popularity bias into embedding updating phase and design a specific debiasing term for each of the four factors accordingly.
\end{itemize}

\subsection{Unbiased Embedding Definition}
As described in Section 3, there are three factors exacerbating the imbalance issue between accumulated positive and negative gradients of the each item: (1) user's high activity interacting with many items; (2) item's high popularity involved in more positive interactions. (3) user's conformity towards other mainstream users. For popular items, its overall gradient is shifted towards the positive gradient direction, making the popular items heavily exposed. The unpopular items in contrast are under-exposed.  The large magnitude of popular item vectors results in higher ranking scores for positive interactions.
Since updating of user and item latent vectors is intertwined in the implicit RS optimization process, we propose to adjust both user and item latent vectors to mitigate popular bias.

Given the item embedding $\boldsymbol{Q_i}$ and user embedding $\boldsymbol{P_u}$ trained using vanilla MF~\cite{koren2009matrix} (or LGN~\cite{he2020lightgcn}) model, we represent the item embedding as:
\begin{equation}
\boldsymbol{Q_i} = \boldsymbol{Q_i^{pop}} + \boldsymbol{Q_i^{int}}
   \label{eq.item}
\end{equation}
where $\boldsymbol{Q_i^{pop}}$ captures the item's popularity and $\boldsymbol{Q_i^{int}}$ captures the item's genuine features. Similarly, we represent a user embedding $\boldsymbol{P_u}$ as:
\begin{equation}
\boldsymbol{P_u} = \|\boldsymbol{P_u^{act}}\|_2(\boldsymbol{P_u^{conf}} + \boldsymbol{P_u^{int}})
\label{eq.user}
\end{equation}
where $\boldsymbol{P_u^{act}}$ captures the users' activeness,  $\boldsymbol{P_u^{conf}}$ signifies the user's conformity to other mainstream users, and $\boldsymbol{P_u^{int}}$ denotes the user's genuine interest.
Notably, taking the product of $\boldsymbol{P_u^{act}}$ and $\boldsymbol{P_u^{conf}}+\boldsymbol{P_u^{int}}$ is justified by the fact that an active user is inclined to have more observed interactions \cite{liu2010personalized}. Formally, the unbiased matching score for a user and an item is:
\begin{equation}
    \hat{y}_{ui} = \boldsymbol{P_u^{int}} \cdot \boldsymbol{Q_i^{int}}
\end{equation}
In the following sections, we will describe how we mitigate $\boldsymbol{P_u^{act}}$ in the training stage and eliminate $\boldsymbol{Q_i^{pop}}$ and $\boldsymbol{P_u^{conf}}$ in the testing stage.

% {\color{red}{unify testing stage and inference stage}}
\subsection{Normalizing User Embeddings in Training}
As an active user will have more observed interactions with items, their accumulated gradients based on these items will likely be larger than those of other less active users. For example, if users $u_1$ and $u_2$ share similar item preferences with $u_1$ being more active, the embedding vectors of $u_1$ and $u_2$ will have similar updating direction but the vector norm of user $u_1$ will be much larger. Such large embedding contributes to the gradient updating of its corresponding items, aggravating the imbalance issue between accumulated positive and negative gradients of items. 

To reduce the bias propagation of highly active users, we normalize each user embedding $\boldsymbol{\hat{P_u}}$ into a unit vector during the training stage:
\begin{equation}
    \boldsymbol{\hat{P}_u} = \boldsymbol{P_u}/\| \boldsymbol{P_u^{act}} \|_2
\end{equation}
where $\boldsymbol{\hat{P}_u}$ is the learned user embedding after normalization. To keep the notation simple, we use $||\boldsymbol{P_u}||$ to denote $||\boldsymbol{P_u^{act}}||_2$. The complete training procedure is shown in Algorithm~\ref{alg:training}.
%combination of user interest and conformity embedding.
% \scalebox{0.75}{
\begin{algorithm}[t]
\caption{Training}
{\small
  \begin{algorithmic}[1] 
    \Require Backbone user encoder $f(\cdot|\Theta_U)$, backbone item encoder $f(\cdot|\Theta_I)$, user set $\mathcal{U}$, item set $\mathcal{I}$, and user-item interactions $Y$.
             \State Initialize the user and item encoder parameters $\Theta_U$ and $\Theta_I$ respectively;
             \State $\boldsymbol{\bar{P}}=0$ and $\boldsymbol{\bar{Q}}=0$
            %   \WHILE {not done}
            \For {$t=1$ to $T$ }
                \State  $Y_t^+ \leftarrow \text{SampleMiniBatch}(Y, m)$ 
                % \Comment{a mini-batch of $m$ positive interactions}
                \State $Y_t^- \leftarrow \text{NegativeSampler}(Y_t^+)$
                \For {$k=1$ to $m$ }
                    \State $u, i, j = Y_{t,k}^+, Y_{t,k}^-$
                    \State $\boldsymbol{P_u}, \boldsymbol{Q_i}, \boldsymbol{Q_j} = f(u|\Theta_U), f(i|\Theta_I), f(j|\Theta_I)$
                    
                    \State $\boldsymbol{\hat{P}_u} = \boldsymbol{P_u}/\| \boldsymbol{P_u} \|_2$
                    
                    \State $\boldsymbol{\bar{P}_u} = \boldsymbol{\bar{P}_u} - lr \cdot (\boldsymbol{Q_i}-\boldsymbol{Q_j})$
                    
                    \State $\boldsymbol{\bar{Q}_i} = \boldsymbol{\bar{Q}_i} - lr \cdot (\boldsymbol{\hat{P}_u})$
                    
                    \State $\boldsymbol{\bar{Q}_j} = \boldsymbol{\bar{Q}_j} - lr \cdot (-\boldsymbol{\hat{P}_u})$
                    
                    \State $\hat{y}_{ui} = \boldsymbol{\hat{P}_u} \cdot \boldsymbol{Q_i}$
                    
                    \State $\hat{y}_{uj} = \boldsymbol{\hat{P}_u} \cdot \boldsymbol{Q_j}$
                    
                    \State $ \mathcal{L}_k = \log \;(\sigma(\hat{y}_{ui}-\hat{y}_{uj}))-\lambda_\Theta ||\Theta||^2$
                \EndFor
                \State $\mathcal{L}_t = \frac{1}{m} \sum^{m}_{k=1} \mathcal{L}_k$ 
                \State $\Theta \leftarrow \Theta - lr \cdot \nabla_\Theta \mathcal{L}(f_\Theta)$
                % \State$L = \sum_{(u,i,j) \in D_s} BPR (\langle P_u/ \|P_u\| , Q_i\rangle,\langle P_u/ \|P_u\|, Q_j\rangle)$\\
            \EndFor
        \State $\boldsymbol{\bar{P}}=\frac{1}{U}\sum^{U}_{u=1} \boldsymbol{\bar{P}_u}$ 
        
        \State
        $\boldsymbol{\bar{Q}}=\frac{1}{I}\sum^{I}_{i=1} \boldsymbol{\bar{Q}_i}$ 
            
    % \ENDWHILE
          \end{algorithmic}
          \label{alg:training}
          }
    \end{algorithm}
    % \vspace{-10px}

\subsection{Adjusting Popularity Bias in Inference}
% \zxy{a short intro of subsection here?}

%In this subsection, we first analyze how the user and item embeddings evolve with item popularity, user conformity, and user activeness in the optimization steps. Then we propose a series of post-hoc interventions to mitigate the corresponding bias in the inference stage.
% \begin{comment}
% \begin{itemize}
% \item Feature embeddings of popular items are overwhelmed by the positive gradients. %As illustrated in \ref{Fig.decomposition},  feature vector of item $i$ consists 
% %
% Compared with a similar but unpopular item, a popular item would be highly exposed since its embedding vector has larger magnitude.
% \item \textcolor{red}{Since an unpopular item has sparse observed interactions with users and being selected as a negative sample in a uniform probability.}  For unpopular items, positive and negative gradients are in relative balance. 
% \textcolor{red}{However, as depicted in Eq.\ref{eq.ranking}, the large magnitude of a popular item will dominate the unpopular one even they share similar property, e.g., similar vector direction.}
% \end{itemize}
% Above all, it is obvious that the magnitude and direction of embedding vector highly influences the recommendation results. In the following, we analysis how the embedding evolves from item popularity, user conformity and user popularity aspects. Corresponding solutions are proposed to mitigate such bias.
% \end{comment}
\subsubsection{Mitigating item popularity influence.}
For a popular item, its positive interactions with users are frequently observed among the training samples, shifting the learned embedding to the positive interaction direction. As depicted in Fig.3, the motivation here is to modify the overall learned item embedding $\overrightarrow{AD}$ to an unbiased item vector $\overrightarrow{AD'}$. There are a number of ways to do so. In this paper, we propose to modify the biased accumulated gradient $\overrightarrow{AD}$ only through its accumulated positive gradients $\overrightarrow{AC}$ and keep $\overrightarrow{AB}$ unchanged.
\begin{figure}[h] 
\centering
\includegraphics[width=0.95\linewidth]{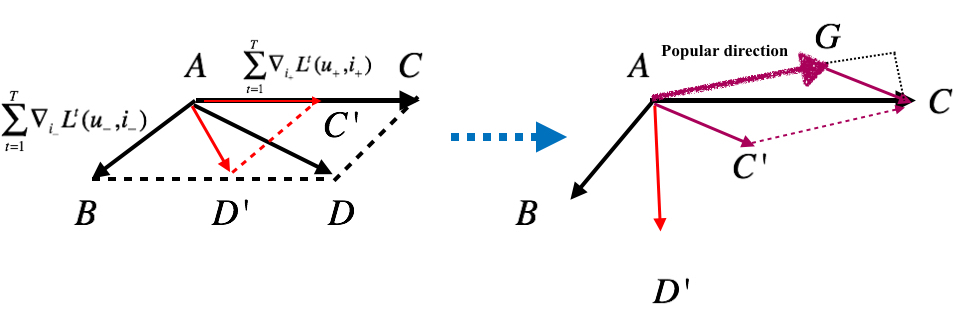} 
\vspace{-10px}
\caption{
$\protect\overrightarrow{AG}$ denotes the popular direction of item embeddings. Our approach adjusts $\protect\overrightarrow{AC}$ to $\protect\overrightarrow{AC'}$ through $\protect\overrightarrow{AG}$, leading to a new modified item embedding $\protect\overrightarrow{AD'}$.
}
\vspace{-5px}
\label{Fig.main3} 
\end{figure}
Nevertheless, the iterative optimization of user/item embedding makes it unclear how much other items and relevant user embeddings contribute to the accumulated positive gradients for each item, worsening the imbalance between $\overrightarrow{AB}$ and $\overrightarrow{AC}$. Hence, the important question here is how to approximate the biased embedding $P_u^{pop}$ in Eq.12. Our analysis in Section 3 suggests that 1) for popular items, their overall accumulated gradient (learned) embeddings ($\overrightarrow{AD}$) are shifted towards its positive gradient direction ($\overrightarrow{AC}$); 
2) Iterative optimization of user and item embedding makes item embeddings are intertwined with other relevant item information. Besides, user conformity and activity makes their item embeddings reflects similar user preference. Thus, popular items share similar positive gradient directions. %}}(EP:??)
%{\color{red} 2) a small fraction of users in observed interaction with a small fraction of popular items makes popular items share similar positive gradient directions.
Based on the two findings,  we use the mean embedding of popular items’ embedding to represent such biased embedding $\boldsymbol{Q^{pop}}$. The direction of such biased vector is named as popular direction $\boldsymbol{\bar{Q}}$.

Thus, we decompose the an item embedding $\boldsymbol{Q_i}$ into a \textit{genuine features embedding} $\boldsymbol{Q_i^{ini}}$ and a projection on the popular direction $\boldsymbol{\bar{Q}}$.  Projection vector here refers to item popularity embedding $\boldsymbol{Q_i^{pop}}$ in Eq.\ref{eq.item}. In the test phase, we adjust the learned item embedding $\boldsymbol{Q_i}$ and derive its genuine representation $\boldsymbol{Q_i^{int}}$ :
\begin{equation}
    \boldsymbol{Q_i^{int}} = \boldsymbol{Q_i} - \boldsymbol{Q_i^{pop}} = \boldsymbol{Q_i} - \alpha_1 \cdot  cos (\boldsymbol{Q_i}, \boldsymbol{\bar{Q}}) \cdot \|\boldsymbol{Q_i}\| \cdot
    \boldsymbol{\bar{Q}}
\end{equation}
where $\alpha_1$ is a scaling parameter which controls the magnitude of the modified positive vector $\overrightarrow{AC'}$. In this paper, we choose the value of $\alpha_1$ based on the validation set.

\subsubsection{Mitigating user conformity}
Different from item embeddings which are updated by both positive and negative gradients, a user embedding is only updated by the gradient $\boldsymbol{Q_i}-\boldsymbol{Q_j}$, in which items $i$ and $j$ appear in observed and non-observed interactions with the user, respectively.
% Although BPR loss makes the assumption that all users are expected to act independently, 
A user embedding is inevitably influenced by the preference of other users during the gradient updates with BPR loss, and thus account for user conformity.

For example, consider a scenario where three users interact with an item $i_1$, while the first user $u$ also interacts with item $i_2$ and $i_3$. Since $i_1$ has multiple observed interactions, it receives the accumulated gradients based on three users. Whenever we update user embedding $\bigtriangledown_u \mathcal{L}(u,i)= \boldsymbol{Q_{i_1}}$. The embedding of $\boldsymbol{Q_{i_1}}$ with large magnitude will increase the magnitude of latent vector of $u$.
%The information flow through the iterative gradient updating, $user->item->user$, makes the user embedding biased.  

The situation here is similar to item popularity bias, we therefore adopt a similar way to extract user's true interest by subtracting the user conformity projection:
\begin{equation}
      \boldsymbol{P_u} = \boldsymbol{P_u} - \alpha_2 \cdot  cos(\boldsymbol{P_u}, \bar{\boldsymbol{P}}) \cdot ||\boldsymbol{P_u}|| \cdot
      \bar{\boldsymbol{P}}
\end{equation}
where $\bar{\boldsymbol{P}}$ denotes the mean or normalized embedding of mainstream users' embedding.
The term $\alpha_2 \cdot  cos(\boldsymbol{P_u}, \bar{\boldsymbol{P}}) \cdot ||\boldsymbol{P_u}|| \cdot
      \bar{\boldsymbol{P}}$ captures the user conformity embedding $\boldsymbol{P_u^{conf}}$ in Eq. 12. $\alpha_2$ is a scaling parameter. We show inference procedure in Algorithm~\ref{alg:inference}.

\begin{algorithm}[t]
{\small
\caption{Adjusting Popularity Bias in Inference}
    \textbf{Input:} Backbone user encoder $f(\cdot|\Theta_U)$, backbone item encoder $f(\cdot|\Theta_I)$, user $u$, item $i$, recorded user conformity $\boldsymbol{\bar{P}}$, and recorded item popularity $\boldsymbol{\bar{Q}}$. \par
    % Backbone user encoder $f_U$, Backbone item encoder $f_I$, User $u$, I set $I$, training dataset $D_s$ 
    \textbf{Output:} $\hat{y}_{ui}$
  \begin{algorithmic}[1] 
            
        \State $\boldsymbol{P_u}, \boldsymbol{Q_i}, \boldsymbol{Q_j} = f(u|\Theta_U), f(i|\Theta_I), f(j|\Theta_I)$
  
         \State $\boldsymbol{Q_i^{int}} = \boldsymbol{Q_i} - \alpha_1 \cdot  cos (\boldsymbol{Q_i}, \boldsymbol{\bar{Q}}) \cdot \|\boldsymbol{Q_i}\| \cdot
         \boldsymbol{\bar{Q}}$
         
         \State $\boldsymbol{P_u^{int}} = \boldsymbol{P_u} - \alpha_2 ,  cos(\boldsymbol{P_u}, \bar{\boldsymbol{P}}) \cdot ||\boldsymbol{P_u}|| \cdot
         \bar{\boldsymbol{P}}$
      
         \State $\hat{y}_{ui} = \boldsymbol{P_u^{int}} \cdot \boldsymbol{Q_i^{int}}$
         
        %  \State $R_u = \{ (i,rank(\| \boldsymbol{Q_i} \| cos(\boldsymbol{P_u},\boldsymbol{Q_i}))) \}_{i \in I}$
    %         \STATE Top-K ranking items
            
    %          \STATE initialization;
    %           \WHILE {Not Converged}
    %             $L = \sum_{(u,i,j) \in D_s} BPR (\langle P_u/ \|P_u\| , Q_i\rangle,\langle P_u/ \|P_u\|, Q_j\rangle)$\\
    % \ENDWHILE
          \end{algorithmic}
          \label{alg:inference}
          }
    \end{algorithm}
    \vspace{-1cm}

\section{Experiments}
\label{sec:expt}
In this section, we conduct experiments to show the effectiveness and efficiency of the proposed framework. Specifically, we aim to answer the following research questions:

\begin{itemize}[leftmargin=*]
    \item \textbf{RQ1}: (a) Does our proposed method outperform state-of-the-art recommendation methods under the non-IID setting? (b) Furthermore, does it generalize well on both BPR and BCE loss? (c) Is it more efficient than state-of-the-art methods? 
    \item \textbf{RQ2}: How do different components (e.g., normalization, user conformity subtrahend, item popularity subtrahend, and these subtrahends’ hyper-parameters\footnote{We show the effect of hyper-parameters $\alpha_1$ and $\alpha_2$ in detail in Appendix.}) affect the result of our method?
    \item \textbf{RQ3}: (a) How well does our method mitigate the popularity bias? (b) How does the performance of our method change with the intervention proportion of test data?
\end{itemize}

\subsection{Experiment Settings}
\begin{table}[t]
\caption{Statistics of Datasets.}
\vspace{-7px}
\centering
% \vspace{-5px}
\resizebox{0.45\textwidth}{!}{
{\small
\begin{tabular}{lccccc}
\toprule
Dataset & User &  Item & Interaction & Sparsity \\
\midrule
Movielens-10M & 37,962 & 4819 & 1,371,473 & 0.007496 \\
Netflix & 32,450 & 8432 & 2,212,690 & 0.008086\\
Adresa & 13,485 & 744 & 116,321 & 0.011594\\
Gowalla & 29,858 & 40,981 & 1,027,370 & 0.000840\\
\bottomrule
\end{tabular}\label{tab::dataset}
}
}
\vspace{-15px}
\end{table}
\textbf{Datasets and Data Preprocessing.} Our experiments utilize four publicly available datasets: 1) Movielens-10M~\footnote{https://grouplens.org/datasets/movielens/} is a dataset with 10M movie ratings. 2) Netflix~\cite{james2007netflix} is from an open competition and it covers film ratings. 3) Adressa~\cite{gulla2017adressa} is a popular dataset for news recommendation. 4) Gowalla~\cite{cho2011gowalla} is a dataset covering location-based user check-in behaviors. We summarize statistics of the datasets in Table~\ref{tab::dataset}. 

To investigate the extent to which the debiased models reduce popularity bias, we replace the conventional evaluation strategy, where the test set consists of IID samples from the observed interactions, with a sampling strategy to construct a debiased test set~\cite{bonner2018causal}. Specifically, we randomly sample interactions with equal probability given to all items to construct the test and validation data.  We then use the remaining data that come with popularity bias as the training data. We follow DICE~\cite{zheng2021disentangling} to process Movielens-10M and Netflix datasets and MACR~\cite{wei2020model} to process Adressa and Gowalla for a fair comparison. More specifically, we sample $30\%$ unbiased interactions regarding items as the test set, $10\%$ as the validation set, and the remaining $60\%$ as training data for Movielens-10M and Netflix. We prepare $10\%$ unbiased data as the test set, another $10\%$ unbiased data as the validation set, and the rest as the training set for Adressa and Gowalla.

\begin{table*}[tbp]
 \centering
    \caption{Overall performance comparison (Recall@20, HR@20, and NDCG@20) of different methods on four datasets. %The best performance methods and the second best ones are shown in boldface and underlined numbers respectively
    }
    \vspace{-7px}
    \label{table:main_results}
    \resizebox{0.88\textwidth}{!}{
    \begin{tabular}{c|c|ccc|ccc|ccc|ccc}
      \toprule
      \multicolumn{2}{c|}{Dataset} & \multicolumn{3}{c|}{Movielens-10M} & \multicolumn{3}{c|}{Netflix} & \multicolumn{3}{c|}{Adressa} & \multicolumn{3}{c}{Gowalla} \\
      \hline
      Model & Method & Recall & HR & NDCG & Recall & HR & NDCG & Recall & HR & NDCG & Recall & HR & NDCG \\
      \midrule
      \multicolumn{1}{c|}{\multirow {7}{*}{MF}} & Original & 0.128	& 0.442	& 0.084 & 0.112 & 0.519 & 0.094 & 0.085 & 0.111 & 0.034 & 0.046 & 0.174 & 0.032 \\
      \multicolumn{1}{c|}{} & IPS & 0.133 & 0.443 & 0.085 & 0.105 & 0.488 & 0.086 & 0.096 & 0.128 & 0.039 & 0.048 &0.174&0.033\\
      \multicolumn{1}{c|}{} & CausE & 0.115 & 0.406 & 0.074  & 0.093 & 0.464 & 0.078 & 0.084 & 0.112 & 0.037 & 0.045 & 0.166 & 0.032 \\
      \multicolumn{1}{c|}{} & DICE & \underline{0.163} & \underline{0.519} & \underline{0.108} & \underline{0.125} & \underline{0.554} & \underline{0.107} & 0.098 & 0.133 & 0.041 & 0.052 & 0.177 & 0.033 \\
      \multicolumn{1}{c|}{} & MACR & 0.138	& 0.469 & 0.089 & 0.119& 0.533 & 0.101  & \underline{0.109} & \underline{0.140} & \underline{0.050} & \underline{0.077} & \underline{0.252} & \underline{0.050} \\
      \multicolumn{1}{c|}{} & \textbf{Ours} & \textbf{0.173} &\textbf{0.536} & \textbf{0.116} & \textbf{0.138} & \textbf{0.575} & \textbf{0.119} & \textbf{0.116} & \textbf{0.147}  & \textbf{0.056} & \textbf{0.081} & \textbf{0.259} & \textbf{0.057} \\\cline{2-14}
    %   &$p$-value &  & &  & &  & &  & &  & &  &  \\
    %   \multicolumn{1}{c|}{} & Improv. & 6.13\% & 3.27\% & 7.40\% & 10.40\% & 3.79\% & 11.21\% & 6.42\% & 5.00\% & 12.00\% & 5.19\% &  2.77\% & 14.00\% \\
    %   \multicolumn{1}{c|}{} & impr\% & 15.7\% & 10.6\% & 17.2\% & 14.5\% & 7.7\% & 16.1\% & 12.1\% & 6.8\% & 12.9\% & 11.7\% & 5.1\% & 12.8\% \\
     %   \multicolumn{1}{c|}{} & impr\% & 35.15\% & 22.12\% & 38.09\% & 23.24\% & 10.78\% & 26.25\% & 36.4\% & 26.59\% & 64.70\% & 76.08\% & 48.85\% & 78.12\% \\ %avg recall: 37.33, 33.87, 51.79
      \midrule
      \multicolumn{1}{c|}{\multirow {7}{*}{LGN}} & Original & 0.137	& 0.462	& 0.089 & 0.102 & 0.490 & 0.087 & 0.098 & 0.123 & 0.040 & 0.045 & 0.172 & 0.032 \\
      \multicolumn{1}{c|}{} & IPS & 0.139 & 0.464 & 0.091 & 0.110 & 0.509 & 0.095 & 0.107 & 0.139 & 0.047 & 0.045 &0.174&0.032\\
      \multicolumn{1}{c|}{} & CausE & 0.102 & 0.372 & 0.063  & 0.083 & 0.428 & 0.067 & 0.082 & 0.115 & 0.037 & 0.046 & 0.173 & 0.033 \\
      \multicolumn{1}{c|}{} & DICE & \underline{0.181} & \underline{0.556} & \underline{0.122} & \underline{0.142} & \underline{0.591} & \underline{0.121} & 0.111 & 0.141 & 0.046 & 0.054 & 0.185 & 0.036 \\
      \multicolumn{1}{c|}{} & MACR & 0.157	& 0.508 & 0.101 & 0.119& 0.531 & 0.094  & \underline{0.127} & \underline{0.158} & \underline{0.052} & \underline{0.077} & \underline{0.254} & \underline{0.051} \\
      \multicolumn{1}{c|}{} & \textbf{Ours} & \textbf{0.186} & \textbf{0.564} & \textbf{0.127} & \textbf{0.145} & \textbf{0.595} & \textbf{0.127} & \textbf{0.134} & \textbf{0.166}  & \textbf{0.067} & \textbf{0.081} & \textbf{0.263} & \textbf{0.053} \\\cline{2-14}
    %   &$p$-value &  & &  & &  & &  & &  & &  &  \\
    %   \multicolumn{1}{c|}{} & Improv. & 2.76\% & 1.43\% & 4.09\% & 2.11\% & 0.67\% & 4.95\% & 5.51\% &  5.06\% & 28.84\% & 5.19\% & 3.54\%  & 3.92\% \\
    %   \multicolumn{1}{c|}{} & Improv. & 46.45\% & 22.07\% & 46.29\% & 42.15\% & 21.42\% & 45.97\% & 36.73\% &  34.95\% & 67.5\% & 80\% & 52.90\%  & 65.62\% \\ avg recall: 51.33, 32.83, 56.34
      \bottomrule
    \end{tabular}}
    % \vspace{-10px}
    %\footnotesize{B.M.: base model; C.M.: causal model.}
    % \\\zxy{is the result significant? try T-test?}
    \vspace{-10px}
\end{table*}

% \noindent\textbf{Evaluation}
% To measure the recommendation performance, we adopt three widely-used evaluation metrics: Recall, Hitting Ratio (HR), which consider whether the relevant items are sorted within the top-K positions, and NDCG that measures the relative orders among positive and negative items in the top-K positions. \textcolor{red}{ In this paper, we set the value of $K$ as}:
% \begin{equation}
% \begin{aligned}
%     DCG_u@K = \sum_{(u,i)\in D}\frac{I(Z_{u,i}\leq K)}{log(\hat{Z_{u,i}}+1)}, \\
%     NDCG_@K = \frac{1}{||U||}\sum_{u \in U}\frac{DCG_@K}{IDCG_@K}.
% \end{aligned}
% \end{equation}
% where $Z_{u,i}$ denotes the rank position of a positive feedback $(u,i)$ and IDCG@K is an ideal DCG@K.

\noindent\textbf{Baselines.} We implement our method over the classical MF and the state-of-the-art LGN to investigate the effectiveness and generalization ability of our proposed approach across different backbone recommendation models. We compare our methods with the following baselines. 
Matrix factorization (\textit{\textbf{MF}}) \cite{koren2009matrix} is a representative collaborative filtering method. \textit{\textbf{LGN}} \cite{he2020lightgcn} is a state-of-the-art collaborative filtering based on light graph convolution  network. \textit{\textbf{IPS}} \cite{chen2020bias} is a widely used re-weighting method based on causal inference . The weighting score, i.e., inverse propensity score is set as the inverse of corresponding item popularity value. It eliminates popularity bias by imposing large weight to unpopular items. \textit{\textbf{CausE}} \cite{bonner2018causal} is a domain adaptation algorithm  that learns a causal embedding from a small uniform dataset. \textit{\textbf{DICE}} \cite{zheng2021disentangling} is the state-of-the art disentangled method for learning causal embedding to cope with popularity bias problem. It designs a framework with causal-specific data to disentangle interest and popularity with pre-defined guidelines. DICE leverage popularity bias in the inference. We use the code provided by authors. \textit{\textbf{MACR}} \cite{wei2020model} addresses popularity bias from a causal-effect view. The counterfactual inference is performed to estimate the direct effect from item properties to the ranking score, which is removed to eliminate the popularity bias.
% \begin{itemize} 
%     \item \textbf{MF}. Matrix factorization is a representative collaborative filtering method \cite{koren2009matrix}.
%     \item \textbf{LGN}. This is a state-of-the-art collaborative filtering based on light graph convolution  network \cite{he2020lightgcn}.
%     \item \textbf{IPS}. This is a widely used re-weighting method based on causal inference \cite{kainz2017improving}. The weighting score, i.e., inverse propensity score is set as the inverse of corresponding item popularity value. It eliminates popularity bias by imposing large weight to unpopular items. 
%     \item \textbf{CausE}. This is a domain adaptation algorithm \cite{bonner2018causal} that learns a causal embedding from a small uniform dataset. 
%     \item \textbf{DICE}. This is the state-of-the art disentangled method \cite{zheng2021disentangling} for learning causal embedding to cope with popularity bias problem. It designs a framework with causal-specific data to disentangle interest and popularity with pre-defined guidelines. Notably, DICE leverage popularity bias in the inference. We use the code provided by authors.
%     \item \textbf{MACR}. This work \cite{wei2020model} firstly addresses popularity bias from a causal-effect view. The counterfactual inference is performed to estimate the direct effect from item properties to the ranking score, which is removed to eliminate the popularity bias.

% \end{itemize}
\noindent\textbf{Implementation Details.} We evaluate the effectiveness performance by three widely-used metrics, i.e., \textit{top-k Recall} (Recall@k), \textit{top-k Hit Ratio} (HR@k), and \textit{top-k Normalized Discounted Cumulative Gain} (NDCG@k), which are used in DICE~\cite{zheng2021disentangling} and \textit{MACR}~\cite{wei2020model}. For MACR on Movielens-10M and Netflix, we use the source code provided by their authors and set the batch size and embedding size as 128 for a fair comparison. For the reported results of the remaining baselines, we refer to DICE and MACR directly. Note that our method is based on two backbones (i.e., MF and LGN). The hyper-parameters of backbones are set following the suggestions from DICE for Movielens-10M and Netflix and MACR for Adressa and Gowalla to guarantee a fair comparison with DICE and MACR. $\alpha_1$ and $\alpha_2$ are determined on validation set based on grid search in the search list formed by taking values from 0 to 2 with a 0.2 increment.
\vspace{-0.1cm}
\subsection{Overall Comparison (RQ1)}
The recommendation performance (i.e., Recall@20, HR@20, and NDCG@20) of different models are presented in Table~\ref{table:main_results}. The methods with best performance are marked with boldface fonts and the second best ones are underlined.
Based on the results, we make the following observations:
% \vspace{-7.59px}
\begin{itemize}[leftmargin=*]
%  \item Our proposed $???$ consistently outperforms baselines with significant improvements with respect to all metrics on four representative datasets.. For example, $???$  makes over 15% improve- ments with respect to NDCG@50 using MF as backbone on Moveilens-10M dataset, and over 20% improvements with re- spect to Recall@20 using GCN as backbone on Netflix dataset.
\item As shown in Table~\ref{table:main_results}, our proposed method consistently outperforms all the baselines on all the datasets w.r.t. all the evaluation metrics. The state-of-the-art baselines DICE and MACR incorporate extra modules to reduce popularity bias, which demands additional prior knowledge and efforts to exploit good architecture by trial and error during the training stage. In contrast, our method only needs to store the aggregated influence of the dominating items and users that is caused by gradient based updates. It efficiently handles popularity bias by performing a post-hoc subtraction calculation in the inference step, which eventually leads to better performance than baselines. 
 
\item Our proposed method is both general and effective. It can easily be incorporated into various backbone RS models such as MF, LGN, and any other general recommendation models. At the same time, it achieves promising performance on debiased test set. Compared to classical MF, our proposed method that uses MF as the backbone model achieves average 37.33\%, 33.87\%, and 51.79\% improvements in Recall@20, HR@20, and NDCG@20, respectively. Similarly, our proposed method boosts the performance of original LGN about 51.33\%, 32.83\%, 56.34\% w.r.t. Recall@20, HR@20, and NDCG@20, respectively. These significant results demonstrate that the method effectively mitigates the popularity bias in the training and test stages. 
% {\color{red} (EP: I thought our method also mitigates bias during inference.)}
%  \item $?$ is a general and efficient framework which can plug in with MF, LGN models and achieve promising results. Without any auxiliary data or complex moduels, ? deconfound popularity bias in the inference stage without any cost. 
% The proposed ? model is sourced from how data is generated and what is the bias factors, thus the proposed framework is independent with backbone recommendation models.

% \zxy{more explanations rather than just numbers?}
 
\item Our proposed method has stronger generalization ability, i.e., it is more robust, as it consistently outperforms others on all four datasets with diverse context and data distribution. While the state-of-the-art methods DICE and MACR achieve comparable performance, their performance may not always be consistent across the different datasets. A possible reason is that DICE and MACR require elaborate configuration of incorporated modules and take much effort to tune the optimal hyper-parameters. Another possible reason to explain the inconsistency of DICE is that DICE depends on two rule-based strategies to obtain negative examples and to employ negative or positive BPR loss to update the model. The strategies are unfortunately sensitive to data distribution.
%  \item In terms of generalization ability on diverse dataset, it is clear that DICE and MACR achieves comparable performance but not always the SOTA across different dataset. As depicted in \cite{zhang2021causal},  DICE over-amplifies the interest for the middle groups that contain sub-popular items, and therefore is sensitive to datset distribution.
\item As to the remaining baselines for mitigating popularity bias, 
CausE has limited improvement over the basic models and sometimes performs even worse. The reason should be that CausE aligns user and item embeddings with a tiny uniform dataset which hardly allows the model to learn enough user preference information. IPS performs poorly in most cases as it simply up-weighs unpopular items with a static score, making the model overly-confident to recommend unpopular items.
% \item As to baselines for mitigating popularity debias, CausE have limited improvement over the basic models and even sometimes perform even worse. The reason should be CausE align user and item embedding with a small uniform data and hardly learn enough user preference information. IPS also performs badly on most of case as it simply upweights unpopular items with a static score, making the model over-confident to \textcolor{red}{sparse unpopular items.}
\end{itemize}
\vspace{-0.3cm}
\subsection{Further Analysis}
% {\color{red} \zxy{divide to multiple subsections, give them more formal names, like Ablation Study, Efficiency Analysis, Case Study, Transferability Analysis, Parameter Analysis, etc}}
In this section, we continue to evaluate our proposed method by answering the remaining research questions:
\begin{figure}[t]
\vspace{-10px}
% \centering
% \RaggedRight
\subfigure[Movielens-10M]{

% \label{traditional}
\includegraphics[width = 4cm]{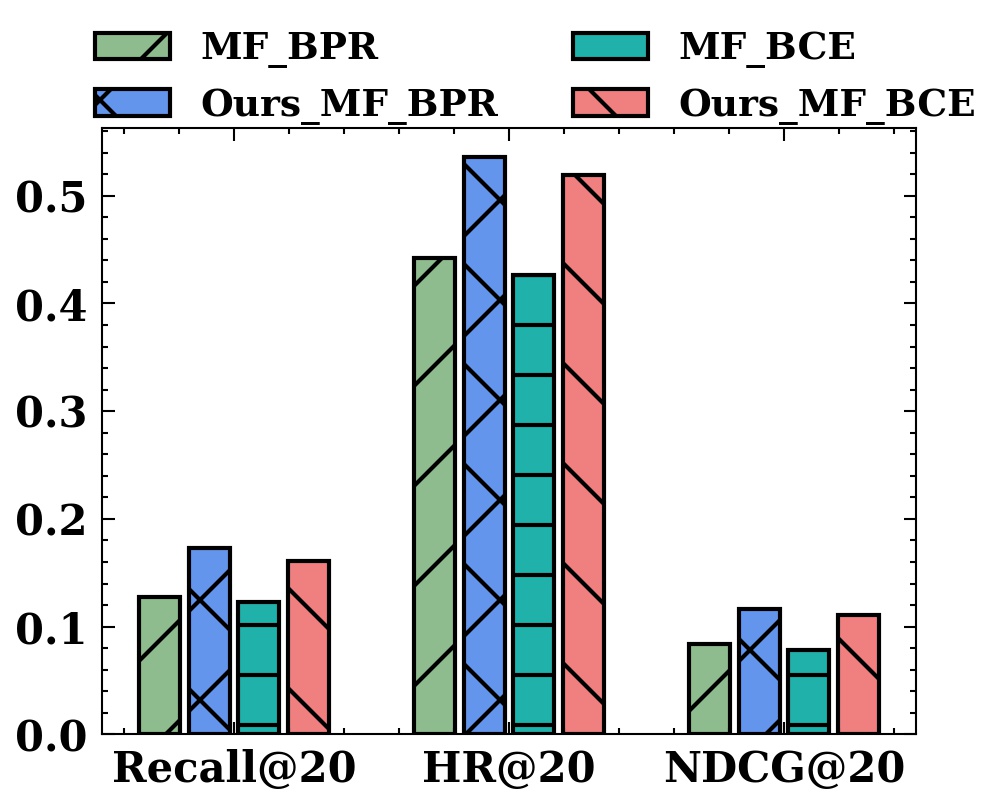}
}
%\subfigure[Recsys that models the popularity bias]{
\subfigure[Netflix]{
\centering
% \RaggedRight
% \label{simple}
\includegraphics[width = 4cm]{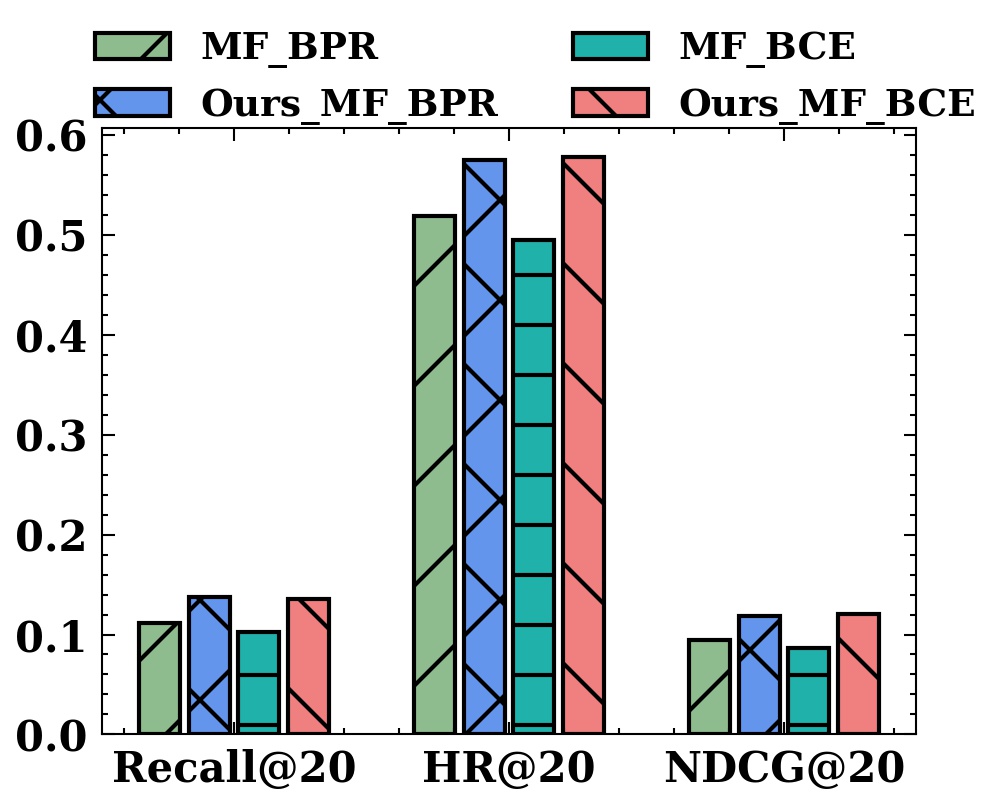}
}
\centering
\vspace{-7px}
\caption{Performance (Recall@20, HR@20, and NDCG@20) of MF and Our Method with BPR and BCE loss on Movielens-10m and Netflix datasets.}
\label{fig:diff_loss}
\vspace{-10px}
\end{figure}

\subsubsection {Does it generalize well on both BPR and BCE loss? (RQ1(b))} Figure~\ref{fig:diff_loss} shows how our method using MF as the backbone performs when applied with BPR or BCE loss on Movielens-10M and Netflix. Overall, subtracting user conformity bias and item popularity bias in the testing stage provides substantially better performance than the models using vanilla user and item embeddings. This observation validates our view that the optimization of BPR and BCE incorporates the user conformity bias and the item popularity bias into user and item embeddings, respectively, making the recommendation results deviating from users' true interest.
The proposed post-hoc design is an elegant and effective way to remove bias.
% Figure \ref{fig:diff_loss} evaluates how our method performs when applying on Movielens-10M and Netflix in terms of BPR and BCE loss. Overall, \textcolor{red}{subtracting conformity bias in the inference time provides substantial higher performance than the Naive loss function. }This observation verifies our insight that optimization of BPR and BCE unifies the conformity bias and interests in user and item embedding, making the recommendation results diverge from users' true interests. \textcolor{blue}{The proposed post-hoc design is an elegant and effective way to remove bias.}

\subsubsection{Is our method more efficient than state-of-the-art baselines? (RQ1(c))}
\begin{table}[t]
    \centering
    \caption{Comparison of computing time for our method versus DICE and MACR. All methods are run on 1 Tesla V100 GPU chip.}
    \vspace{-5px}
    \resizebox{0.36\textwidth}{!}{
    \begin{tabular}{c|c|cc}
    \toprule
    \multicolumn{2}{c}{} &\multicolumn{2}{c}{Seconds per epoch} \\ 
    \cmidrule(r{1em}l{1em}){3-4} 
      Model & Method  & Movielens-10M & Netflix \\
       \midrule
     \multicolumn{1}{c|}{\multirow {4}{*}{MF}} &Original    & 47 & 77   \\
    %  \midrule
    \multicolumn{1}{c|}{} &DICE & 124 ($2.63$x)& 207 ($2.68$x)\\
     \multicolumn{1}{c|}{}&MACR & 57 ($1.21$x)&  87($1.29$x)\\
      \multicolumn{1}{c|}{}&Ours & 51 ($1.08$x)& 78 ($1.01$x)\\
       \midrule
     \multicolumn{1}{c|}{\multirow {4}{*}{LGN}} &Original    & 126 & 201   \\
    %  \midrule
    \multicolumn{1}{c|}{} &DICE & 307 ($2.43$x)& 454 ($2.25$x)\\
     \multicolumn{1}{c|}{}&MACR & 304 ($2.41$x)& 447 ($2.22$x)\\
      \multicolumn{1}{c|}{}&Ours & 129 ($1.02$x)& 203 ($1.01$x)\\
     
     \bottomrule
     
    \end{tabular}
    }
    
    \label{tab:efficiency}
    \vspace{-10px}
\end{table}
We investigate the efficiency of our proposed method by comparing the running time on the same device against the two state-of-the-art methods, i.e., DICE and MACR, with MF and LGN backbones. As shown in Table~\ref{tab:efficiency}, we observe that our method achieves comparable running time with classical MF and LGN methods, and it is significantly more efficient than DICE and MACR. With special causal learning design and direct supervision on disentanglement, DICE takes twice more time than our proposed method due to the additional cost of parameter updating. MACR includes additional modules to reconstruct user and item frequencies, and is therefore less efficient. Another interesting side observation is that DICE always suffers from excessive computation time compared with MACR.  We conjecture that the additional two rule-based strategies of DICE are partially responsible for the excess time consumed.  

% We conduct ablation study to investigate the computation time for ? with current two state-of-the-art methods, i.e., DICE and MACR, in terms of MF and LGN backbones. We can observe that, ? consumes comparable running time with vanilla MF and LGN and it is significantly more efficient than DICE and MACR. \textcolor{red}{MACR in LGN???}
% With special causal learning design and direct supervision on disentanglement, DICE takes more than twice times as the proposed method due to the cost of additional parameter updating. MACR designs extra modules to reconstruct user and item frequency, and therefore, it is inevitable to be less efficient in terms of computation time. An interesting phenomenon is that DICE always suffers from excessive computation time compared with the MACR model. We conjecture that it is because DICE does not explore the cause factors behind conformity and interests explicitly from the optimization perspective. They design the disentanglement supervision based on prior principles which require splitting a dataset to compute a discrepancy loss term. The deficiency of theoretical analysis makes such design choice inefficient in computation time and inaccurate when the data distribution violates their assumption.

%\input{Tables/different_loss}

\subsubsection{Is it necessary to normalize user embeddings and is it desired to subtract user conformity and item popularity term? (RQ2)} 
\begin{table}[tbp]
\caption{Ablation study (Recall@20, HR@20, and NDCG@20) of our method on Movielens-10M and Adressa datasets.}
\centering
\vspace{-5px}
\scalebox{0.86}{\footnotesize
\begin{tabular}{c|c|ccc|ccc}
\toprule
\multicolumn{2}{c}{Datasets}&\multicolumn{3}{c}{Movielens-10M}& \multicolumn{3}{c}{Adressa} \\
\cmidrule(r{1em}l{1em}){3-5}  \cmidrule(r{1em}l{1em}){6-8}
Model & Method & Recall &  HR & NDCG & Recall &  HR & NDCG \\
\midrule
\multicolumn{1}{c|}{\multirow {4}{*}{MF}} & Ours & 0.173 & 0.536 & 0.116 & 0.116 & 0.147 & 0.056 \\
\multicolumn{1}{c|}{}& w/o norm  & 0.164 & 0.517 & 0.109 & 0.098 & 0.125 & 0.044  \\ %a1=0.2, a2=0.6
\multicolumn{1}{c|}{}& $\alpha_1=0$  & 0.153 & 0.497 & 0.101 & 0.088 & 0.113 & 0.043  \\
\multicolumn{1}{c|}{}& $\alpha_2=0$  & 0.160 & 0.520 & 0.111 & 0.089 & 0.133 & 0.049 \\
\midrule
\multicolumn{1}{c|}{\multirow {4}{*}{LGN}} & Ours & 0.186 & 0.564 & 0.127 & 0.134 & 0.166 & 0.067 \\
\multicolumn{1}{c|}{}& w/o norm  & 0.171 & 0.532 & 0.114 & 0.122 & 0.148 & 0.052  \\
\multicolumn{1}{c|}{}& $\alpha_1=0$  & 0.154 & 0.498 & 0.101 & 0.119 & 0.151 & 0.061  \\
\multicolumn{1}{c|}{}& $\alpha_2=0$  & 0.175 & 0.540 & 0.117 & 0.113 & 0.144 & 0.051  \\
\bottomrule
\end{tabular}\label{tab::ablation}
}
\vspace{-10px}
\end{table}
We conduct an ablation study with our method using MF and LGN backbones on Movielens-10M and Adressa datasets to analyze the contribution of different components. Specifically, we compare our method with its three special variants: our method without (w/o) user norm (i.e., MF w/o norm and LGM w/o norm), where normalization on user embedding is removed; our method with $\alpha_1 = 0$, where we just simply drop the user conformity embedding term in the inference; our method with $\alpha_2 = 0$ that drops the item popularity embedding term.

The results in Table~\ref{tab::ablation} show that removing any component would lead to worse performance. It indicates that all three components contribute to improving the model performance under non-IID circumstances. Omitting user embedding normalization would reduce accuracy in all metrics on both Movielens-10M and Adressa datasets. This validates the usefulness of normalizing user embedding magnitudes in the training stage, i.e., promoting items with fewer interactions benefits recommendation performance. It is clear that compared with the variant without item popularity bias, the model performs much worse when removing the user conformity bias in most cases. One possible reason is that user has more impact than item for these two datasets.

\subsubsection{How does our method mitigate the popularity bias? (RQ3(a))}
\begin{figure}[ttbp]
% \centering
% \RaggedRight
\subfigure[MF]{
% \label{traditional}
\includegraphics[width = 4cm]{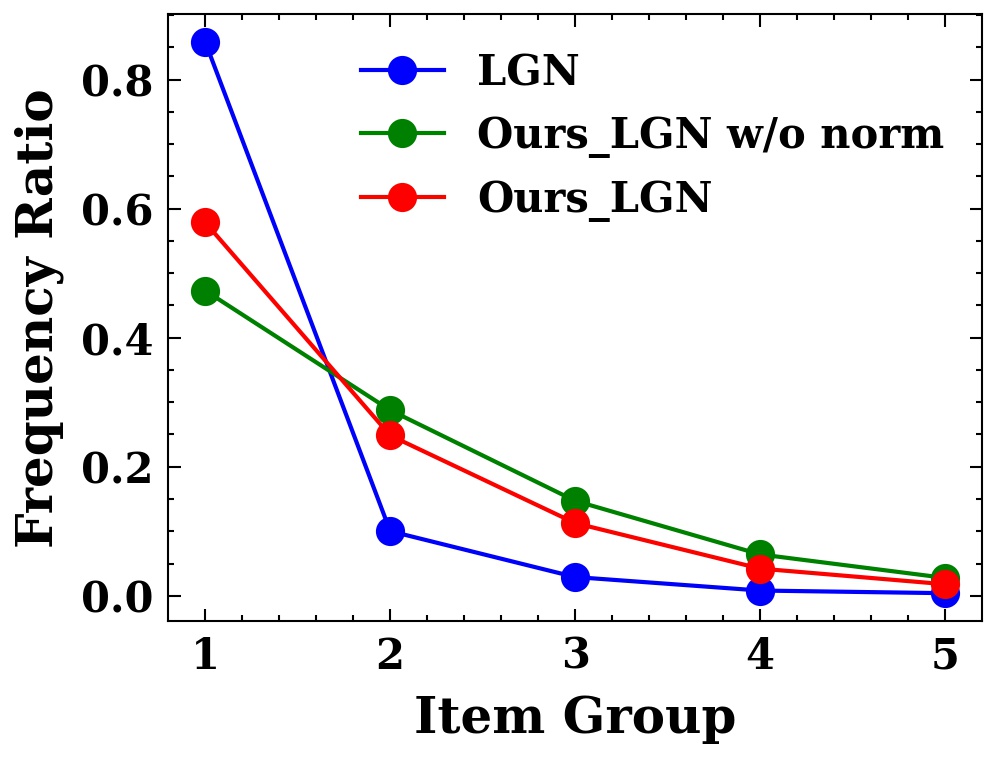}
}
%\subfigure[Recsys that models the popularity bias]{
\subfigure[LGN]{
\centering
% \RaggedRight
% \label{simple}
\includegraphics[width = 4cm]{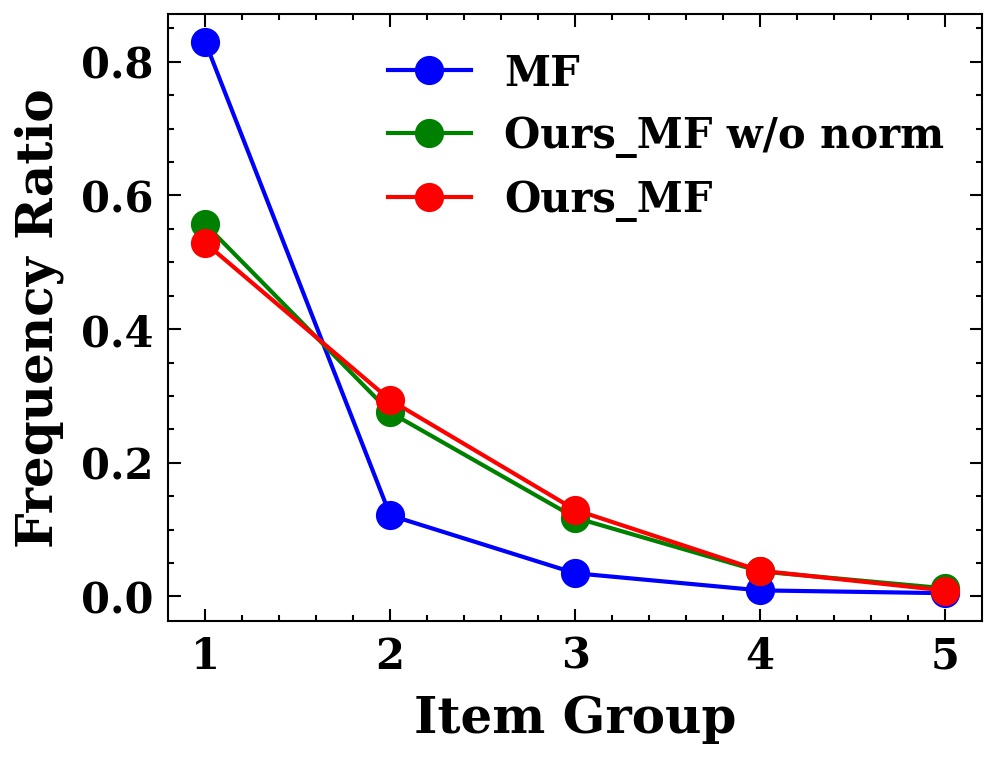}
}
\centering
\vspace{-7px}
\caption{Recommended frequency of different item groups by MF (LGN), Our\_MF (Our\_LGN) w/o norm, and Our\_MF (Our\_LGN) on Movielens-10M.}
\label{fig:freq_ratio}
\vspace{-10px}
\end{figure}

\begin{figure}[ttbp]
% \centering
% \RaggedRight
\subfigure[MF]{

% \label{traditional}
\includegraphics[width = 4cm]{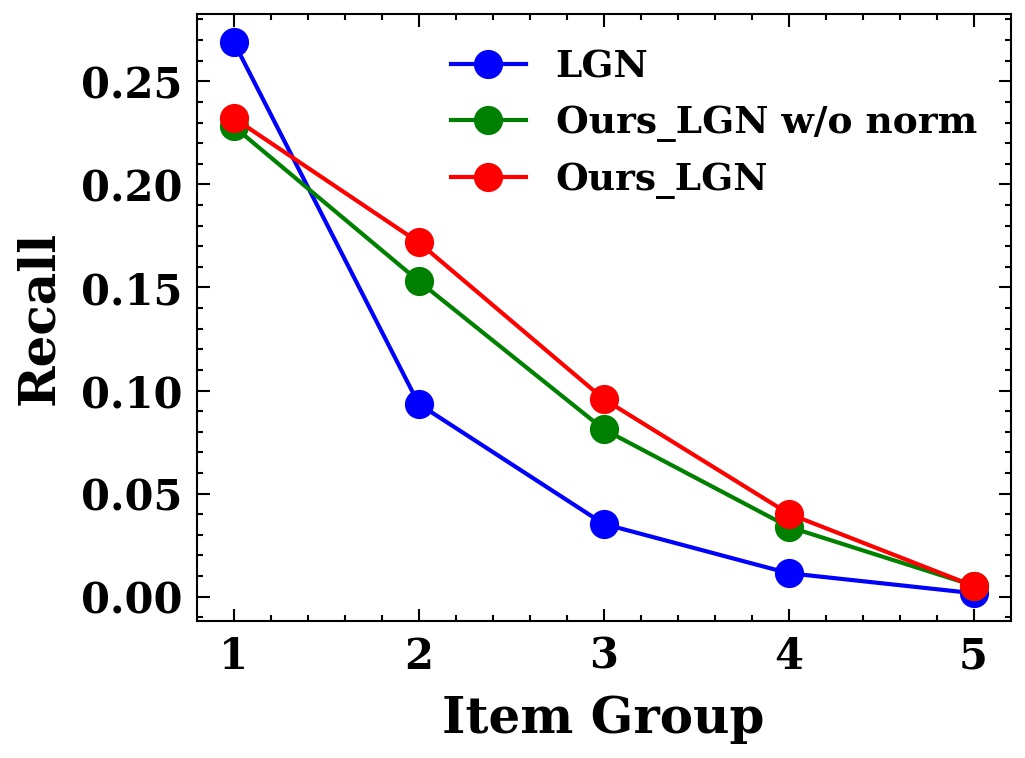}
}
%\subfigure[Recsys that models the popularity bias]{
\subfigure[LGN]{
\centering
% \RaggedRight
% \label{simple}
\includegraphics[width = 4cm]{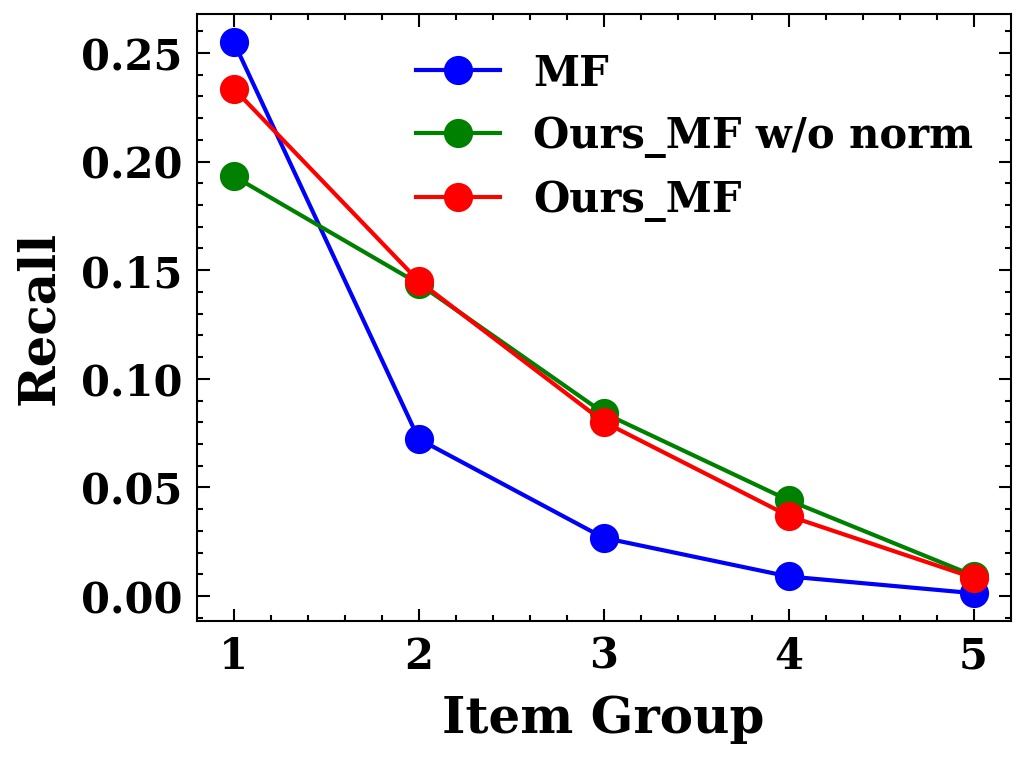}
}
\centering
\vspace{-7px}
\caption{Performance (Recall@20) of MF (LGN), Our\_MF (Our\_LGN) w/o norm, and Our\_MF (Our\_LGN) in different item groups on Movielens-10M. }
\label{fig:group_recall}

\vspace{-10px}
\end{figure}

In this study, we investigate whether our method can remove the effect of popularity bias by the methods, i.e., MF (LGN), Our\_MF (Our\_LGN) w/o norm, and Our\_MF (Our\_LGN) on Movielens-10M. For fine-grained debias analysis, we first sort items in inverse popularity order. We then divide items into five groups, each of the first four groups accounts for $5\%$ of items, and we put the rest into the last group. Group 1 represents the most popular item group, and Group 5 represents the least popular item group. For each group, we calculate the average recommended frequency and recall@20. We show the recommended frequency of each group by methods in Figure \ref{fig:freq_ratio} and recall in Figure ~\ref{fig:group_recall}.
% We make a brief analysis below.

Conventional recommendation methods have the most significant recommended frequency and perform best in group 1 but yield a significant performance drop in group 2 and further drops in other groups. It indicates that conventional recommender systems tend to recommend popular items to unrelated users due to user conformity and popularity bias.
In contrast, conventional recommender systems with our proposed post-hoc method have less recommended frequency and performance in the most popular group but achieve a significantly higher recommended frequency and better recommendation performance from group 2 to group 4. It shows that our proposed method effectively reduces user conformity and item popularity bias.

In group 5, all methods achieve similar recommended frequency and performance. The possible reason is that these least popular items rarely appear in the training set so that the representation of these items cannot be learned effectively. In some cases, our method achieves less recommended frequency but higher recall than our method w/o norm. It means that our method can still more accurately recommend items to users.

% \vspace{-0.5cm}
\begin{figure}[t!]

\centering 
\includegraphics[width=4.5cm]{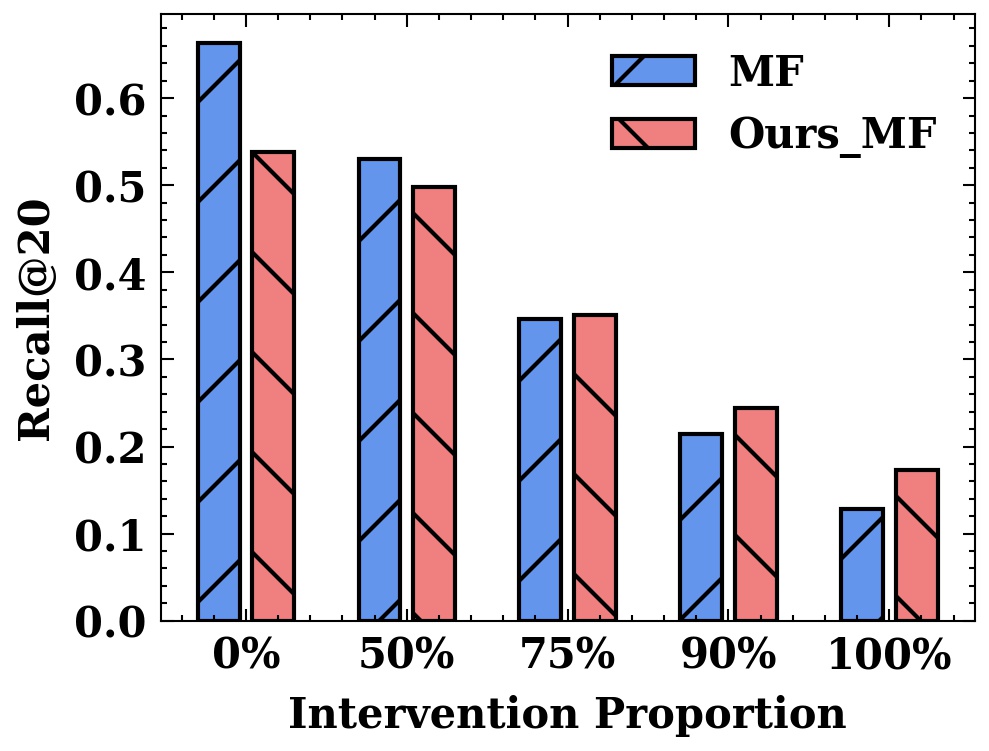} 
\vspace{-10px}
\caption{Performance of MF and Ours\_MF with different proportion of intervened test data.}
\label{Fig:imbalance_ratio} 
\vspace{-10px}
\end{figure}

\subsubsection{How does our method's performance change along with intervention proportion of test data? (RQ3(b))}
In previous experiments, all methods are tested on the 100\% intervened test data. This study investigates how methods perform as the proportion of intervened test data changes. We evaluate MF and our method (MF) with five different proportions of intervened test data, i.e., 0\%, 50\%, 75\%, 90\%, 100\%. 0\% means training and test data have the same distribution. 100\% means all test data are intervened data. As shown in Figure 7, our method starts to outperform MF on 75\% of intervened test data and outperforms most significantly under 100\% intervened test data, which verifies the effectiveness of our method under non-IID circumstances.

% \begin{figure}[t!]

% \centering 
% \includegraphics[width=4.5cm]{Figures/imbalance_ratios.jpg} 
% \vspace{-10px}
% \caption{Performance of MF and Ours\_MF with different proportion of intervened test data. \zxy{text too small} \wl{revised, and the figure for LGN is to be added}}
% \label{Fig:imbalance_ratio} 
% \vspace{-10px}
% \end{figure}

% \zxy{a case sutdy of active v.s. inactive users?}

% \subsubsection{Visualization of Embeddings}

%\subsubsection{Running Time}

% \subsection{A Close Look}
% In this section, we dive into the proposed method and study its unique characteristics.

% \input{Tables/number_ui}
% \subsubsection{Statistics of Popular Items and Active Users}

% For supporting the analysis of popularity bias from a gradient perspective. We take the training data of our experiment dataset Movielens-10M as an example, and statistics is shown in Table~\ref{tab:v}.
% We sort items by their number of interactions in descending order and select sorted items until the number of interactions of selected items up to 80\% of total interactions in the training set (in fact, less than 20\% items account for more than 80\% interactions in many real scenarios). We assume this selected items as popular items and the remaining items as unpopular items. In the same way, we can get active users and inactive users. 

% \subsubsection{Magnitude of item and user embeddings.}

% Here, we show the magnitude of item and user embeddings in inverse popularity and activity order, respectively. It justifies the popularity bias mentioned in Section 3. 

\subsubsection{Effect of the hyper-parameters of subtrahends}
\begin{figure}[H] 
\centering 
\includegraphics[width=7.5cm]{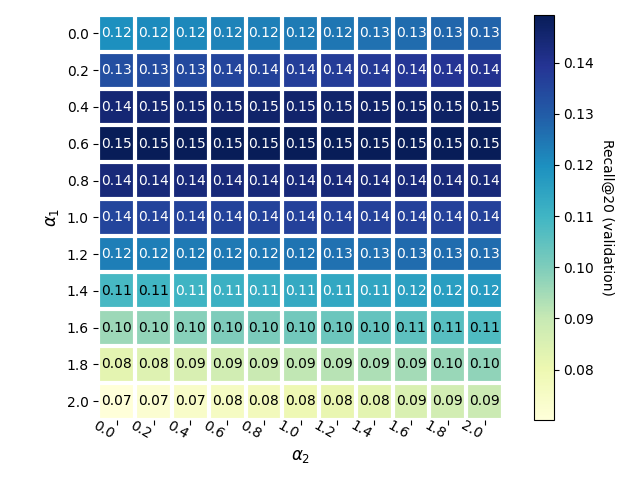} 
\vspace{-10px}
\caption{Effect of $\alpha_1$ and $\alpha_2$ in our method (MF) for mitigating the user conformity bias and the item popularity bias respectively.}
\label{Fig:hyper-parameter} 
\vspace{-10px}
\end{figure}

Our post-hoc method has two hyper-parameters $\alpha_1$ and $\alpha_2$ for the user conformity subtrahend and the item popularity subtrahend respectively. We experiment with our method using MF as the backbone on Movielens-10M to investigate their effect. Figure~\ref{Fig:hyper-parameter} shows how our method's performance (recall@20) changes as $\alpha_1$ and $\alpha_2$  change from 0 to 2 over the interval. 

As we can see, the model performs well when $\alpha_1$ is between 0.4 to 0.6 and performs increasingly better while increasing $\alpha_2$ in most cases. It suggests that the appropriate amount of mitigating user conformity bias and item popularity bias benefits the recommendation system under non-IID circumstances. Large $\alpha_1$ yields a drop in performance while large $\alpha_2$ improves the recommendation performance, which indicates that excessive user conformity debiasing hurts the performance. One possible reason is that user normalization overly constrains the magnitude of user embeddings.

% \input{Tables/number_ui}

% \begin{figure}[H] 
% \centering
% \includegraphics[width=1.0\linewidth]{Figures/ana_ui_2.pdf} 
% \vspace{-20px}
% \caption{Illustrations for the effect of popularity on the norm of user and item latent vectors on Movielens-10M. (a) Magnitude (L2 norm) of the latent vectors of users in inverse item popularity order; (d) Magnitude of the latent vectors of items in inverse user activity order. }
% \label{fig:analysis_1} 
% \vspace{-10px}
% \end{figure}

\section{Related Work}
In recent years, recommendation systems have achieved great success due to the merit of deep learning methods, most of which aim at the development of novel machine learning models to fit user behavior data.
When it comes to real-world scenarios, the distribution of user and item interaction is inevitably long-tailed, making popularity bias a long-standing challenge \cite{ren2021fair,joachims2017unbiased,ren2022semi}. 
%Popular items are overly exposed in recommendations and such bias promotes misleading rich-get-richer dynamics.

The first core idea in mitigating popularity bias is to balance the data distribution such that popular and unpopular items are equally important in the training process.
Inverse propensity weighting (IPW) \cite{austin2015moving,ren2021cross} is widely used to assign weights inversely proportional to item popularity in the empirical risk of a RS model \cite{schnabel2016recommendations}. While such reweighting methods achieve significant improvement for under-represented items, they come with a high potential risk of overfitting and suffer from high variance \cite{guo2020survey,yao2020survey}. 
%Note that all algorithms proposed by these works evaluate popularity bias by comparing how often items are recommended without regard for the ground truth of user-item matching.
%
Recent work \cite{sacharidis2020building, zhao2018zero, xiao2017fairness,zhao2022exploring} studies this problem from a fairness perspective \cite{francez2012fairness,zhao2020balancing}. and enforce the algorithm to produce a fair allocation of exposure based on merit. 
Other strategies following the first pipeline is to transfer the knowledge learned from popular groups to unpopular groups \cite{liu2020general}. 
%What to transfer and how to transfer are the fundamental questions to answer under this methods \cite{torrey2010transfer, pan2009survey}.
Such methods often introduce extra noise to unpopular items and modify the initial dataset distribution. Recent work \cite{zhang2021causal} investigates the bias from the perspective of causal inference and proposes a counterfactual reasoning method using Do-calculus~\cite{pearl2009causality}. They propose a different causal graph based on prior knowledge, however, the essence is to re-balance the distribution for popular and unpopular groups with a causal language. 

Another line of research is to preserve specific user and item information intentionally, with the goal to make the unpopular items more representative \cite{li2021leave}. With auxiliary text information, \cite{li2021leave} an autoencoder layer \cite{zhao2021graphsmote,kingma2013auto} is added while learning user and item representations with text-based Convolutional Neural Networks. Latest work \cite{bonner2018causal} regards a personalized recommendation system as a treatment policy utility maximization problem and proposes a domain adaptation method named CausE.
\cite{chen2021autodebias} propose a meta-learning strategy and transfer unbiased data representation from a uniform logged data. However, the availability and high-quality of auxiliary dataset is the bottleneck for such approaches. 
%To ease the burden of extra dataset, \cite{} design a user and an item module to reconstruct the user and item frequency at the cost of extra parameter updating and computational time.
%they either significantly increase the parameters or require a complicated training strategy

The third categories try to solve this problem through ranking adjustment \cite{abdollahpouri2017controlling, zhao2020semi,ren2018tracking}. These approaches result in a trade-off between the recommendation accuracy and the coverage of unpopular items.
They typically suffer from accuracy drop due to give consideration to the long-tail in a brute manner. 

Different from previous method, in this paper we do not focus on balancing the distribution of popular and unpopular items groups or making unpopular items more representative. We provide a new insight that unbalancing between the positive and negative gradients of the same item incurs the popularity bias. Some related works \cite{rendle2014improving} discuss the gradient vanish of unpopular items from a gradient perspective in implicit recommendation system. However, they still focus on the unbalanced distribution between popular and unpopular item groups, which actually fall into the former three categories.
\section{Conclusion}

In this paper, we propose a simple, efficient, and general framework to mitigate popularity bias from a gradient (first-order information) perspective. In our approach, we record accumulated gradients of users and items as popularity bias during the training stage. Then we address popularity bias by gradient-level calibration during the testing stage as a post-hoc debiasing strategy. We implement our proposed method over MF and LGN on four real-world datasets. Extensive experiments demonstrate that our approach outperforms competitive baselines, e.g.,  DICE and MACR. As part of the future work, we will consider applying our approach to more complex recommendation tasks, such as sequential recommendation and conversational recommendation.

%\zxy{Please use Grammarly to check grammar mistakes and typos of the whole paper!!}
\section{Acknowledgement}
This research was partially supported by the National Science Foundation (NSF) via the grant numbers: 2040950, 2006889, 2045567.

\bibliographystyle{IEEEtran}
\bibliography{ICDM}

\newpage
% \appendix
% \input{Sections/Appendix}
\end{document}